\documentclass[fleqn,usenatbib]{mnras}

\usepackage{newtxtext,newtxmath}

\usepackage[T1]{fontenc}

\DeclareRobustCommand{\VAN}[3]{#2}
\let\VANthebibliography\thebibliography
\def\thebibliography{\DeclareRobustCommand{\VAN}[3]{##3}\VANthebibliography}

\usepackage{graphicx}	
\usepackage{amsmath}	
\usepackage{mathtools}

\usepackage{booktabs}
\usepackage{float}

\usepackage{orcidlink}

\usepackage{subfigure}
\usepackage{caption}

\title[Distinct quenching origins in high-z galaxy clusters]{Distinct origins of environmentally quenched galaxies in the core and outer virialised regions of massive clusters at $0.8<z<1.5$}

\author[G. Hewitt et al.]{Guillaume Hewitt$^{1,2,3}$\thanks{E-mail: guillaume.hewitt@nottingham.ac.uk}\orcidlink{0009-0006-7827-007X}, Florian Sarron$^{4}$\orcidlink{0000-0001-8376-0360}, Michael L. Balogh$^{2,3}$\orcidlink{0000-0003-4849-9536}, Gregory Rudnick$^{5}$\orcidlink{0000-0001-5851-1856}, Yannick Bah\'e$^{1,6}$\orcidlink{0000-0002-3196-5126},
\newauthor Devontae C. Baxter$^{7}$\orcidlink{0000-0002-8209-2783}, Gianluca Castignani$^{8}$\orcidlink{0000-0001-6831-0687}, Pierluigi Cerulo$^{9}$\orcidlink{0000-0003-0703-3123}, M. C. Cooper$^{10}$\orcidlink{0000-0003-1371-6019},
\newauthor Ricardo Demarco$^{11}$\orcidlink{0000-0003-3921-2177}, Adit H. Edward$^{12,2,3}$, Rose A. Finn$^{13}$\orcidlink{0000-0001-8518-4862}, Ben Forrest$^{14}$\orcidlink{0000-0001-6003-0541}, Adam Muzzin$^{12}$\orcidlink{0000-0002-9330-9108},
\newauthor Julie Nantais$^{15}$\orcidlink{0000-0002-7356-0629}, Benedetta Vulcani$^{16}$\orcidlink{0000-0003-0980-1499}, Gillian Wilson$^{17}$\orcidlink{0000-0002-6572-7089}, Dennis Zaritsky$^{18}$\orcidlink{0000-0002-5177-727X}
\\
$^{1}$ School of Physics and Astronomy, University of Nottingham, Nottingham, NG7 2RD, U.K \\
$^{2}$ Department of Physics and Astronomy, University of Waterloo, Waterloo, Ontario N2L 3G1, Canada \\
$^{3}$ Waterloo Centre for Astrophysics, University of Waterloo, Waterloo, Ontario, N2L3G1, Canada \\
$^{4}$ IRAP, Institut de Recherche en Astrophysique et Plan\'etologie, Universit\'e de Toulouse, UPS-OMP, CNRS, CNES, 14 avenue E. Belin, F-31400 Toulouse, France \\
$^{5}$ University of Kansas, Department of Physics and Astronomy, 1251 Wescoe Hall Drive, Room 1082, Lawrence, KS 66049, USA \\
$^{6}$ Laboratory of Astrophysics, \'{E}cole Polytechnique F\'{e}d\'{e}rale de Lausanne (EPFL), Observatoire de Sauverny, 1290 Versoix, Switzerland\\
$^{7}$ Department of Astronomy and Astrophysics, University of California, San Diego, 9500 Gilman Dr, La Jolla, CA 92093, USA \\
$^{8}$ INAF-Osservatorio di Astrofisica e Scienza dello Spazio di Bologna, Via Piero Gobetti 93/3, 40129 Bologna, Italy \\
$^{9}$ Departamento de Ingenier\'ia Inform\'atica y Ciencias de la Computaci\'on, Universidad de Concepci\'on, Chile \\
$^{10}$ Center for Cosmology, Department of Physics and Astronomy, University of California, Irvine, 4129 Reines Hall, Irvine, CA 92697, USA \\
$^{11}$ Institute of Astrophysics, Facultad de Ciencias Exactas, Universidad Andr\'es Bello, Sede Concepci\'on, Talcahuano, Chile \\
$^{12}$ Department of Physics and Astronomy, York University, 4700 Keele St. Toronto, Ontario, M3J 1P3, Canada \\
$^{13}$ Department of Physics and Astronomy, Siena College, 515 Loudon Road, Loudonville, NY 12211, USA \\
$^{14}$ Department of Physics and Astronomy, University of California, Davis, One Shields Avenue, Davis, CA, 95616, USA \\
$^{15}$ Instituto de Astrof\'isica, Departamento de F\'isica y Astronom\'ia, Universidad Andres Bello, Fernandez Concha 700, Las Condes 7591538, Santiago, Chile \\
$^{16}$ INAF-Osservatorio astronomico di Padova, Vicolo Osservatorio 5, I-35122 Padova, Italy \\
$^{17}$ Department of Physics, University of California, Merced, 5200 North Lake Road, Merced, CA 92543, USA \\
$^{18}$ Steward Observatory, University of Arizona, 933 North Cherry Avenue, Tucson, AZ 85721-0065, USA
}

\date{Accepted XXX. Received YYY; in original form ZZZ}

\pubyear{2025}

\begin{document}
\label{firstpage}
\pagerange{\pageref{firstpage}--\pageref{lastpage}}
\maketitle

\begin{abstract}
High-redshift ($z\sim1$) galaxy clusters are the domain where environmental quenching mechanisms are expected to emerge as important factors in the evolution of the quiescent galaxy population. Uncovering these initially subtle effects requires exploring multiple dependencies of quenching across the cluster environment, and through time. We analyse the stellar-mass functions (SMFs) of 17 galaxy clusters within the GOGREEN and GCLASS surveys between $0.8<z<1.5$, and with $\log{(M/{\rm{M_\odot}})}>9.5$. The data are fit simultaneously with a Bayesian model that allows the Schechter function parameters of the quiescent and star-forming populations to vary smoothly with cluster-centric radius and redshift. The model also fits the radial galaxy number density profile of each population, allowing the global quenched fraction to be parameterised as a function of redshift and cluster velocity dispersion. We find the star-forming SMF to not depend on radius or redshift. For the quiescent population however, there is $\sim2\sigma$ evidence for a radial dependence. Outside the cluster core ($R>0.3\,R_{\rm200}$), the quenched fraction above $\log{(M/{\rm{M_\odot}})}=9.5$ is $\sim40{\rm\;per\,cent}$, and the quiescent SMF is similar in shape to the star-forming field. In contrast, the cluster core has an elevated quenched fraction ($\sim70{\rm\;per\,cent}$), and a quiescent SMF similar in shape to the quiescent field population. We explore contributions of `early mass-quenching' and mass-independent `environmental-quenching' models in each of these radial regimes. The core is well-described primarily by early mass-quenching, which we interpret as accelerated quenching of massive galaxies in protoclusters, possibly through merger-driven feedback mechanisms. The non-core is better described through mass-independent, environmental-quenching of the infalling field population.
 
\end{abstract}
\defcitealias{Remco2020}{vdB20}
\begin{keywords}
galaxies: clusters: general -- galaxies: evolution -- galaxies: general -- galaxies: high-redshift.
\end{keywords}



\section{Introduction}\label{sec:Intro}
The observed bimodality of galaxies in colour \citep{Strateva2001,Blanton2003,Baldry2004,Mateus2006}, star formation rate \citep{Brinchmann2004,Elbaz2007,Brammer2011,McGee2011,RenziniPeng2015}, and morphology \citep{Cattaneo2006,Driver2006,Mignoli2009}, has established that the process of transformation from actively star-forming to passive (quenching) is an important feature of late-time galaxy evolution. This bimodality is present from the local Universe through to cosmic noon, and is strongly dependent on the galaxy’s stellar mass \citep{Kauffmann2003,Baldry2004,Bundy2006,Ilbert2010,Ilbert2013}, and its local environment \citep{Kauffmann2004,Balogh2004a,Balogh2004b,Baldry2006,Muzzin2012}. Foundational results over the past few decades have found that these dependencies may be separable in the local Universe, suggesting a corresponding bimodality in the processes that cause quenching \citep{Baldry2006,Peng2010,Muzzin2012,Kovac2014}. Mass-dependent quenching is attributed to internal processes, such as active galactic nuclei (AGN) \citep{SilkRees1998,Croton2006,Fabian2012}, or stellar feedback mechanisms \citep{DekelSilk1986,WhiteFrenk1991,Hopkins2012}. Environmental-dependent quenching is attributed to external processes, with mechanisms that are caused by gravitational interactions of neighbouring objects \citep{Moore1996,Moore1998,Bialas2015,Smith2015}, or the interaction with the intra-group/cluster medium (ICM) of the large-scale structure they are a part of \citep{GunnGott1972,Larson1980,BaloghNaMo2000,Peng2015}.

Quenching efficiency is also observed to be a function of redshift, and the apparent separability between mass and environment no longer holds at $z\gg0$ \citep{Balogh2016,Kawinwanichakij2017,Papovich2018,Pintos-Castro2019,Remco2020}. In particular, the work of \citet{Balogh2016} and \citet{Remco2020} (hereafter \citetalias{Remco2020}) analysed massive galaxy clusters at higher redshifts ($0.8 < z < 1.4$), with both finding a strong mass dependence for the environmental quenching component. This indicates that massive galaxies are preferentially quenched in clusters \emph{in addition to} the mass-dependency of field mass-quenching. 

Since the star-forming and quiescent galaxies have distinct stellar mass function (SMF) shapes, an environment-dependent quenching mechanism should, in general, result in SMFs that also depend on environment, such that the low-mass slope of the quiescent SMF should be steeper in dense environments, as it is locally \citep[e.g.][]{Peng2010}. However, \citetalias{Remco2020} found no evidence for this. This led them to suggest an alternative environmental quenching mechanism in high-redshift clusters that has a mass-dependence that precisely mimics that of mass-quenching.  Indeed, by analysing the fraction of quenched galaxies as a function of both cluster-centric radius and redshift, \citet{Baxter2022} and \citet{Baxter2023} showed that the data are consistent with strongly mass-dependent environmental quenching models. While these models lead to SMFs that are qualitatively similar to the observed cluster SMFs \citep{Baxter2022}, there are still important differences, including a low-mass slope for the quiescent galaxies that is steeper than observed. Recent work by \citet{Gully2025} used additional observations of four of these clusters, probing lower stellar masses. In this subset, they found a significant deviation from the field SMFs for the quiescent population, with the cluster low-mass slopes being steeper, as expected in environmental-quenching models. 

One possibility for why \citetalias{Remco2020} did not observe the signature of environment-quenching in the shape of the stellar mass functions is that the analysis was done on the full cluster ensemble, which spans a wide range in halo mass and redshift. Galaxy populations are known to vary significantly with cluster-centric distance \citep[e.g.][]{Pintos-Castro2019,Baxter2023}, and while the clusters have a range of virial radii, the conclusions were based on an analysis of all galaxies within $1$ Mpc of the cluster core. Moreover, the clusters have mass completeness limits that vary from system to system by an order of magnitude. This warrants a closer look at how the data constrains the SMF parameters as a function of these properties.

In this paper we fit the SMFs of 17 $0.8<z<1.5$ galaxy clusters from GOGREEN \citep{Balogh2021} in a Bayesian manner that allows the shape parameters of each population (star-forming and quiescent) to independently vary with cluster-centric radius and redshift. With indication of further dependencies from previous literature results \citep{Lubin2002,Poggianti2006,Nantais2016,Nantais2017,Oh2018}, we also constrain the radial profiles of the quiescent and star-forming galaxies to find the quenched fraction, and fit parameters that describe how it varies as a function of redshift and cluster velocity dispersion.

The structure of the paper is as follows: In Section \ref{sec:Data} we describe the spectroscopic and photometric data used in our cluster sample, with the galaxy classification selection and sample weights applied, as well as the data reduction and characterisation of an additional cluster novelly added to the sample. Section \ref{sec:LikelihoodChapter} presents the formalism of our likelihood model, and its validation. We present our results in Section \ref{sec:Results}. In Section \ref{sec:Discussion} we discuss our findings in the context of previous literature and explore potential environmental quenching pathways. We summarise the work and give our conclusions in Section \ref{sec:Conclusions}.

In this work we adopt the $\Lambda$CDM cosmology parameters with $\Omega_m=0.3$, $\Omega_\Lambda=0.7$, and $H_0=70$ km s$^{-1}$ Mpc$^{-1}$. All magnitudes stated are in the absolute bolometric (AB) magnitude system. Any uncertainties presented are given at the 1-$\sigma$ level, unless otherwise stated.

\section{Cluster sample data}\label{sec:Data}

The data used in this study come from two surveys, the Gemini Observations of Galaxies in Rich Early ENvironments (GOGREEN) survey \citep{Balogh2017} and the Gemini CLuster Astrophysics Spectroscopic Survey (GCLASS) \citep{Muzzin2012}. Together, the surveys include 26 galaxy clusters, spanning a redshift range of $0.87<z<1.46$ \citep{Balogh2021}. In this study, we restrict our analysis to the fourteen clusters drawn from the \emph{Spitzer} Adaptation of the Red Cluster Sequence (SpARCS) survey \citep{Muzzin2009,Wilson2009,Demarco2010}, and the three clusters from the South Pole Telescope (SPT) cluster survey \citep{Brodwin2010,Foley2011,Stalder2013}.  The cluster centre is defined as the location of the brightest cluster galaxy (BCG) of each cluster. There are two exceptions to this, SpARCS-1047 and SpARCS-1051, which have BCGs that are offset from the centre of the cluster galaxy distribution, and so their centres are chosen to match the positions given in \citet{Remco2013}. The virial radius ($R_{200c}$)  for 14 of the clusters were taken from \citet{Biviano2021}, while the values for SpARCS-0219 and SpARCS-1034 were taken from \citet{McNab2021}. The calculation for SpARCS-1033 is described in \S~\ref{sec:1033}.

All clusters in the sample include deep, multiwavelength photometry using a variety of instruments, as well as substantial multiobject spectroscopy using the Gemini GMOS spectrographs. The majority of the clusters have imaging in the \emph{ugriz} bands, the $Y$/$J_1$ and $K_s$ bands, and all the IRAC channels, at depths that are slightly shallower than COSMOS/UltraVISTA. The photometric redshifts were fit using the \texttt{EAZY} code \citep{EAZY}, and validated with available spectroscopic redshifts.  After applying a small bias correction to the photometric redshifts to ensure a mean $\Delta z$ of zero, the scatter is $\sigma_z=0.048$, with 4.1 per cent outside the `outlier' limit of $|\Delta z|>0.15$, $\left( \text{where } \Delta z = \frac{z_{\rm phot}-z_{\rm spec}}{1+z_{\rm spec}}\right)$ \citepalias{Remco2020}. The high spectroscopic completeness ($\sim40$ per cent) allows the calculation of an accurate statistical membership correction to apply to the photometric sample (see \S~\ref{sec:MCF}). The stellar masses for each galaxy were determined by fitting the observed SEDs using the \texttt{FAST} code \citep{FAST}, where the SED was normalised to match the total $K_s$-band magnitude. This assumed a \citet{Chabrier2003} initial mass function (IMF), solar metallicity, and the dust law from \citet{Calzetti2000}.

The properties of all clusters in the sample are summarised in Table \ref{tab:Cluster_values}.

\subsection{GOGREEN/GCLASS DR2 and SpARCS-1033}
The data catalogues were taken from a preliminary version of the second public data release (DR2) of the GOGREEN and GCLASS surveys. The main difference of DR2 compared with DR1 \citep{Balogh2021} is the addition of the cluster SpARCS-1033 to the photometric sample.  The spectroscopy and redshifts for this cluster were included in DR1, but as the K-band imaging had not yet been completed, the cluster was frequently excluded from previous literature which analysed the DR1 sample.  The image reduction, object catalogues, and characterisation of this cluster are described in \S~\ref{sec:1033}.  The data for this cluster is then incorporated into the data release structure following that of DR1.

In addition, the DR2 release includes a number of minor additions and corrections.  These are described in Balogh et al. (in prep).

\subsubsection{SpARCS-1033}\label{sec:1033}
As part of the GOGREEN survey, images of the SpARCS-1033 field were obtained with Subaru-HSC ($y,z$), Subaru-SuprimeCam ($g,i,r$) and CFHT WIRCam ($J, K$).  Data were reduced as described in \citetalias{Remco2020}, and PSF-matched photometry was done by convolving all images to $1.07$ arcseconds, which corresponds to the worst image quality among all the ground-based images for this cluster. $K_s$-band selected catalogues were obtained from the WIRCam image, with a limiting magnitude of 23.9 AB, corresponding to a stellar mass limit of $\log{M/M_\odot}=9.74 \pm$0.03. (Appendix~\ref{appex:MagCompl}). As with the other clusters, photometric redshifts were measured using \texttt{EAZY}, while the other physical parameters, like stellar mass, were measured using \texttt{FAST}.  This catalogue was then cross-matched with the spectroscopic redshift catalogue from \citet{Balogh2021}.

\begin{figure}
\centering
\includegraphics[width=0.99\linewidth]{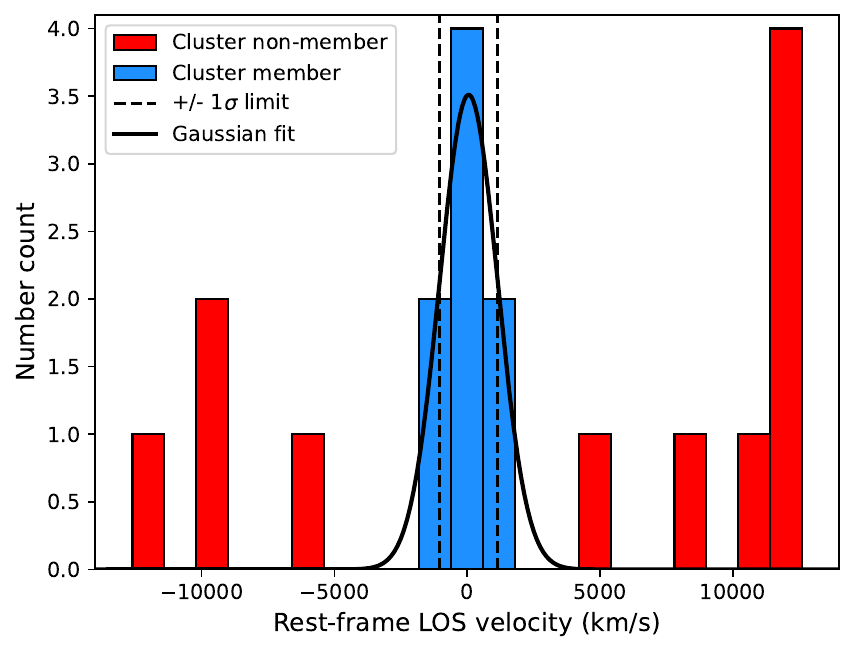}
\caption{Histogram of the line-of-sight velocity of spectroscopically confirmed cluster members of SpARCS-1033 and non-members within 1 Mpc. The dashed lines show the velocity dispersion range of the cluster (see \S~\ref{sec:1033}), and the solid line is a Gaussian distribution with this width.}
\label{fig:1033_veldist}
\end{figure}

The line-of-sight velocity dispersion of SpARCS-1033 was calculated using the standard deviation of eight spectroscopic cluster members, defined as being within a projected distance of 1 Mpc and having a redshift differential with the cluster centre of less than 0.02. This gives a velocity dispersion of 1090 $\pm$ 290 km s$^{-1}$. The distribution of cluster members and non-members in velocity space is shown in Figure \ref{fig:1033_veldist}.

\begin{figure}
\centering
\includegraphics[width=0.99\linewidth]{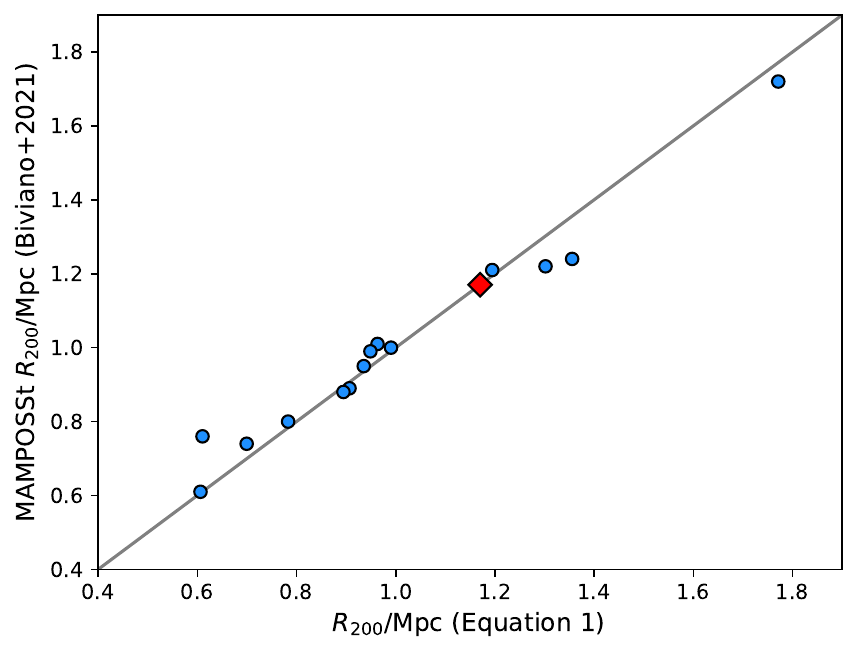}
\caption{Comparison of $R_{200}$ values calculated using MAMPOSSt, and their equivalent $R_{200}$ values calculated using Eq. \ref{eq:r200_eq}. The value of the $A$ parameter (see \S~\ref{sec:1033}) was fit to minimise the difference between the two methods of calculation ($A=2.91 \pm 1.76$). The blue points correspond to the 14 clusters analysed in \citet{Biviano2021}. The red diamond shows the radius of SpARCS-1033 predicted from Eq. \ref{eq:r200_eq} (note that it is not included in the fit). The grey line shows the one-to-one agreement between the two methods.}
\label{fig:r200_fit}
\end{figure}

As previously mentioned, the virial radii for the rest of the clusters were taken from \citet{Biviano2021}, who calculated them using the MAMPOSSt method \citep{Mamon2013}. Since there are few spectroscopic members for SpARCS-1033, using the full dynamical analysis in \citet{Biviano2021} is not well justified. Instead, we compute $R_{200}$ for SpARCS-1033 from the relation
\begin{equation}\label{eq:r200_eq}
    R_{200}=\sqrt{\frac{A}{100}}\frac{\sigma_{\rm LOS}}{H(z)}
\end{equation}
where $\sigma_{\rm LOS}$ is the line-of-sight velocity dispersion, and where the normalisation constant $A$ was determined by calculating a linear fit between Equation~\ref{eq:r200_eq} and the \citet{Biviano2021} values for the rest of the clusters in the sample. This fit is shown in Figure \ref{fig:r200_fit}, with the values having small residuals about the one-to-one line of agreement. Applying it to SpARCS-1033 yields a virial radius of 1.17 $\pm$ 0.35 Mpc, corresponding to a virial mass of 9.33$\times$10$^{14}M_{\odot}$.

\subsection{Galaxy Classification}\label{sec:Classification}
After excluding stellar objects (those with the \texttt{Star} flag set to 1), the sample of seventeen cluster fields consists of 57,528 galaxies.

Potential cluster members are identified by the difference in redshift relative to the tabulated cluster average, $\Delta z_{i} = \frac{z_{i}-z_\mathrm{clust}}{1+z_{i}}$. Following \citetalias{Remco2020}, spectroscopic cluster members are defined as those with $|\Delta z_{\rm spec}|\leq 0.02$, which includes galaxies within 2-3 times the typical cluster velocity dispersion. Candidate photometric cluster members are defined as those with $|\Delta z_{\rm phot}|\leq 0.08$, chosen so the number of false-positives is comparable to the number of false-negatives, minimizing the membership correction (see \S~\ref{sec:MCF}). We also limit the analysis to galaxies within $R_{200}$ of each cluster. This results in a sample of 2,780 potential members. A weight to correct for projection effects is described in \S~\ref{sec:MCF}.  

\begin{figure}
\centering
\includegraphics[width=0.99\linewidth]{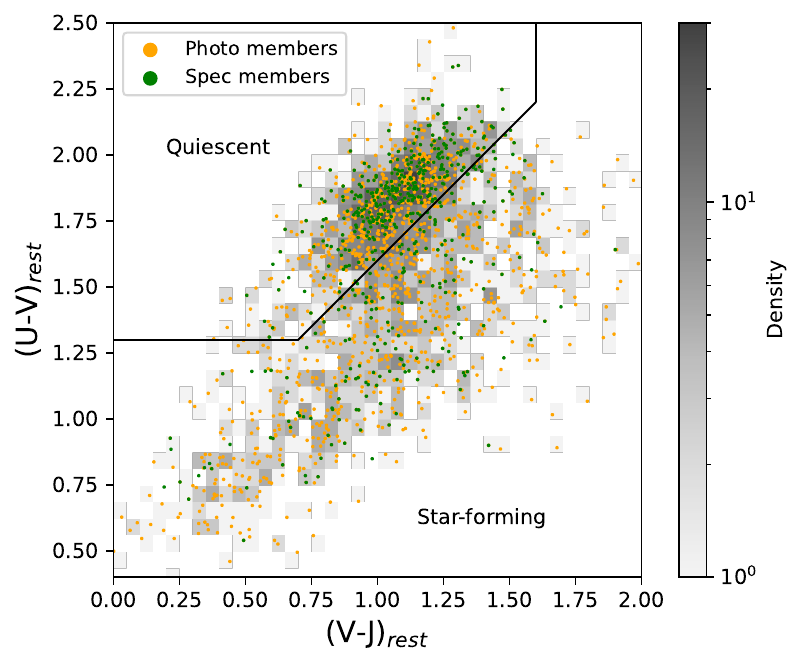}
\caption{Rest-frame \emph{U-V} vs. \emph{V-J} colours for all cluster members with $R<R_{200}$ and $M>M_\mathrm{lim}$, respective to each cluster. The orange points are photometric members and the green points are spectroscopic members, with the gradient background showing the point density, regardless of member type. The black line indicates the distinction between star-forming and quiescent galaxies, given in Eq. \ref{eq:UVJ_cut}.}
\label{fig:UVJ_cut}
\end{figure}

We classify galaxies as star-forming or quiescent based on the rest-frame \emph{UVJ} colours as described in \citetalias{Remco2020}. Specifically, quiescent galaxies are defined as:
\begin{equation}\label{eq:UVJ_cut}
    U-V\,>\,1.3\;\; \cap \;\; V-J\, <\, 1.6\;\; \cap \;\;U-V\, >\, 0.60+(V-J)
\end{equation}
The rest-frame colour distribution and classification is shown in Figure \ref{fig:UVJ_cut}, where the rest-frame colours are plotted for galaxies with stellar masses of $M>M_{\rm lim}$ (the limiting stellar mass for each cluster, based on limiting $K_s$-band magnitudes - see \S~\ref{sec:Compl_weight}), and projected distances of $R<R_{200}$ from their cluster centre. Alternative colour boundaries were also considered \citep[e.g.][]{Muzzin2013}, but these did not have a significant effect on our results and did not affect our conclusions.

\begin{table*}
\centering
\caption{\label{tab:Cluster_values} Overview of the 17 GOGREEN and GCLASS clusters and their properties.}
\begin{tabular}{@{}llllllllll@{}}
\toprule
\textbf{Cluster Name} & \textbf{RA}$^\textbf{BCG}_\textbf{J2000}$ & \textbf{Dec}$^\textbf{BCG}_\textbf{J2000}$ & \textbf{Redshift} & \textbf{$\sigma$}$_\textbf{los}$$^\text{(b)}$ & $R_\textbf{200}$$^\text{(c)}$ & \textcolor{black}{$M_\textbf{200}$} & $K_\textbf{s,lim}$$^\text{(d)}$ & $M_\textbf{*,lim}$ $^\text{(d)}$ & $\textbf{N}_\textbf{SF}$/$\textbf{N}_\textbf{Q}$$^\text{(e)}$ \\
 &  &  & & \textbf{[km s$^{-1}$]} & \textbf{[Mpc]} & \textcolor{black}{[\textbf{log$_{10}$} \textbf{M}$_\odot$]} & \textbf{[mag$_{\text{AB}}$]} & [\textbf{log$_{10}$} \textbf{M}$_\odot$] \\
\midrule
SpARCS-0034 & 8.67512 & -43.1315 & 0.867 & 700 $\pm$ 150 & 0.61 & 13.8 & 21.53 & 10.42 & 15 / 13  \\
SpARCS-0036 & 9.18756 & -44.1805 & 0.869 & 750 $\pm$ 90 & 1.21 & 14.7 & 22.11 & 10.53 & 37 / 34 \\
SpARCS-1613 & 243.311 & 56.825 & 0.871 & 1350 $\pm$ 100 & 1.72 & 15.2 & 22.55 & 9.97 & 114 / 113 \\
SpARCS-1047$^{(a)}$ & 161.8893 & 57.68706 & 0.956 & 660 $\pm$ 120 & 0.99 & 14.5 & 22.68 & 10.17 & 39 / 39 \\
SpARCS-0215 & 33.84996 & -3.72561 & 1.004 & 640 $\pm$ 130 & 0.89 & 14.4 & 21.73 & 10.45 & 31 / 23 \\
SpARCS-1051$^{(a)}$ & 162.79675 & 58.30075 & 1.035 & 689 $\pm$ 40 & 0.95 & 14.5 & 24.17 & 9.35 & 44 / 35 \\
SPT-0546 & 86.6403 & -53.761 & 1.0669 & 977 $\pm$ 70 & 1.22 & 14.8 & 23.47 & 9.64 & 104 / 96 \\
SPT-2106 & 316.5191 & -58.7411 & 1.1307 & 1055 $\pm$ 85 & 1.24 & 14.9 & 23.19 & 9.79 & 95 / 122 \\
SpARCS-1616 & 244.1722 & 55.7534 & 1.156 & 782 $\pm$ 40 & 1.00 & 14.6 & 23.76 & 9.59 & 51 / 78 \\
SpARCS-1634 & 248.6542 & 40.3637 & 1.177 & 715 $\pm$ 40 & 0.88 & 14.5 & 24.01 & 9.50 & 35 / 46 \\
SpARCS-1638 & 249.7152 & 40.64525 & 1.196 & 564 $\pm$ 30 & 0.74 & 14.3 & 23.94 & 9.54 & 29 / 24 \\
SPT-0205 & 31.451 & -58.4803 & 1.3227 & 678 $\pm$ 60 & 0.80 & 14.1 & 23.25 & 9.90 & 24 / 44 \\
SpARCS-0219 & 34.9315 & -5.5249 & 1.325 & 810 $\pm$ 80 & 0.79 & 14.4 & 23.27 & 9.90 & 29 / 16 \\
SpARCS-0035 & 8.957 & -43.206604 & 1.335 & 840 $\pm$ 50 & 1.01 & 14.7 & 23.81 & 9.70 & 89 / 36 \\
SpARCS-0335 & 53.7648 & -29.48217 & 1.368 & 542 $\pm$ 30 & 0.76 & 14.4 & 22.91 & 10.07 & 32 / 13 \\
SpARCS-1034 & 158.706 & 58.30919 & 1.385 & 250 $\pm$ 30 & 0.24 & 12.9 & 24.22 & 9.55 & 7 / 13 \\
SpARCS-1033 & 158.3565 & 57.89 & 1.461 & 1090 $\pm$ 290 & 1.17 & 15.0 & 23.89 & 9.74 & 58 / 28 \\
\bottomrule
\end{tabular}
\begin{minipage}{\linewidth}
    \hspace{0.5cm} \\
    \textbf{\emph{Notes.}} $^a$Clusters with offset BCGs, overridden with positions given in \citet{Remco2013}.\\
    $^b$Line-of-sight velocity dispersion and their 1-$\sigma$ uncertainties for each cluster. All values from \citet{Balogh2021}, with exception of SpARCS-1033.\\
    $^c$Virial radii of each cluster. Values are all from \citet{Biviano2021}, using the $r_{200,Mc}$ profile combination, with the exception of SpARCS-0219 and 1034, which were taken from \citet{McNab2021}, and SpARCS-1033 (calculated in this work).\\
    $^d$The 80 per cent completeness limits, given in magnitude and stellar mass, respectively. Values are taken from \citetalias{Remco2020}, with the exception of SpARCS-0034, 0036, 1613, 1047, and 0215, which are taken from \citet{Remco2013}, and SpARCS-1033 (calculated in this work). All values are based on the $K_s$ detection band, with the exception of SpARCS-0036, which uses the deeper $J$-band. The stellar mass limits correspond to the quiescent population of each cluster. For the star-forming population, the limit extends down by 0.2 dex due to their smaller mass-to-light ratio, in accordance with \citetalias{Remco2020}. Again, to follow consistency with \citetalias{Remco2020}, we also set the global stellar-mass limit to be $>10^{9.5}M_\odot$. \\ 
    $^e$The number of star-forming and quiescent galaxies in each cluster, after all the cuts described in \S~\ref{sec:Classification} and \S~\ref{sec:Weights}. are applied.
\end{minipage}
\end{table*}

\begin{figure}
\centering
\includegraphics[width=0.99\linewidth]{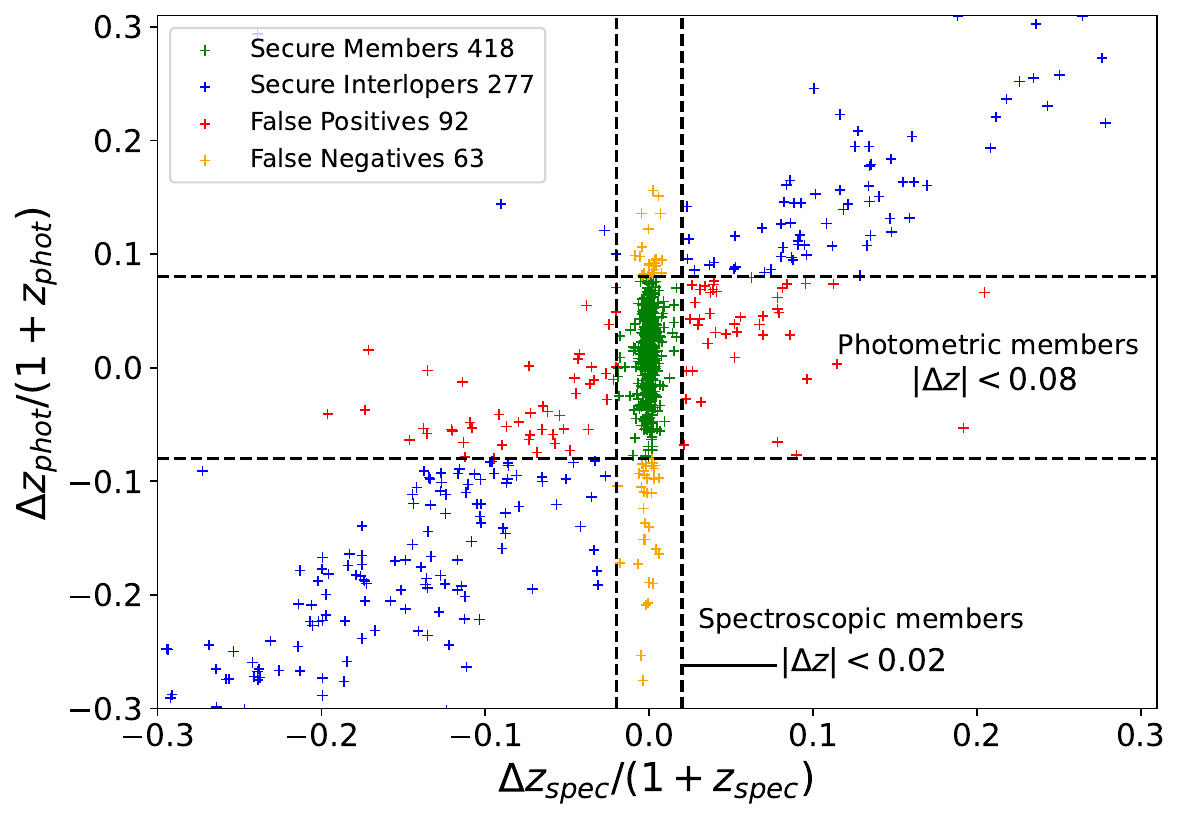}
\caption{For the spectroscopic sample, we compare their photometric and spectroscopic redshift offsets with respect to their clusters. Green and yellow points represent galaxies that are identified as spectroscopically-confirmed cluster members, as shown by the vertical dashed lines. The horizontal dashed lines represent the broader selection of candidate cluster members based on photometric redshifts. Red points are false positives: galaxies that would be identified as a candidate member based on their photometric redshift, but are spectroscopically confirmed non-members.  Blue points are secure interlopers, outside the membership criteria based on either photometric or spectroscopic information. Only galaxies with a stellar mass $M>M_\mathrm{lim}$ and within $R<R_{200}$ of their respective cluster are shown.}
\label{fig:photspec_dist}
\end{figure}

\subsection{Sample Weights} \label{sec:Weights}
\subsubsection{Membership Correction Factor}\label{sec:MCF}
For the galaxies with only photometric redshifts, we must correct for the effects of interlopers and missing members.  We do this following the procedure described in \citetalias{Remco2020}, using the $\sim 40$ per cent of the sample that has spectroscopic redshifts to calculate a correction factor. For the spectroscopic subset we define `true' cluster members as those with $|\Delta z_{\rm spec}|\leq 0.02$, and identify galaxies as either ``secure member'',``secure interloper'',  ``false negative'' or ``false positive'' depending on the value of their $|\Delta z_{\rm phot}|$. The distribution of galaxies with spectra and their designations can be seen in Figure \ref{fig:photspec_dist}. A membership weight for each photometric member \emph{i} is then determined by identifying the five nearest galaxies in stellar mass and projected cluster-centric radius with spectroscopic redshifts. From these five galaxies we compute a correction factor for the corresponding photometric candidate, as: 
\begin{equation}\label{eq:MCF}
    \text{Corr}_i = \frac{\mathcal{N}(\text{secure member}) + \mathcal{N}(\text{false negative})}{\mathcal{N}(\text{secure member}) + \mathcal{N}(\text{false positive})}
\end{equation}
where the $\mathcal{N}(X)$ terms are the number of each respective group out of the five neighbours. In this way, we use the spectroscopic sample to calculate a weight for every photometric galaxy, which depends on its stellar mass and cluster-centric radius. Galaxies that are identified as spectroscopic members have their membership factor set to unity, regardless of their photometric classification. 

\subsubsection{Completeness Weight}\label{sec:Compl_weight}
We include a weight to correct for incompleteness in the photometric sample, based on the $K_s$-band measurements of each galaxy. The weight is structured as the inverse of the completeness, and since the sample is constrained to where recovery is >80 per cent (see Appendix \S~\ref{appex:MagCompl}), the completeness weight is constrained from 1.0-1.25. The completeness value for a specific galaxy depends on its $K_s$-band magnitude, and the limiting $K_s$ magnitude ($K_{s,lim}$) of the image of its parent cluster. These limiting magnitude values of each cluster are given in Table \ref{tab:Cluster_values}, with the majority of the values and the associated limiting cluster stellar mass ($M_{*,lim}$) matching those from \citetalias{Remco2020}, excluding the value for SpARCS-1033, which we calculate using the same procedure (see Appendix \S~\ref{appex:MagCompl}).

\subsubsection{Final sample and weights}
The overall weight applied to each galaxy is the product of the membership correction factor and the completeness weight, shown in this relation:
\begin{equation}\label{eq:Weights}
    w_i=\text{Corr}_i \times \frac{1}{\text{Compl}(K_{s,i})}
\end{equation}

The analysis in this paper is based on unbinned data, using the weight of Equation~\ref{eq:Weights}.  When these unbinned fits are compared with binned data (as in Figures \ref{fig:Blue_SMFs} and \ref{fig:Red_SMFs}), an additional correction is needed to account for the variation of limiting cluster stellar mass across the sample clusters \textcolor{black}{(see Column 9 in Table \ref{tab:Cluster_values})}. Following \citetalias{Remco2020}, this is done by applying an additional weight (to the binned data only):
\begin{equation}
    w_i=
    \sum\limits_{cl}\lambda_{cl}\left(\sum\limits_{\substack{cl \\ M_{*,i}>M_{*,lim,cl}}}\lambda_{cl}\right)^{-1},
\end{equation}
where $\lambda$ is the cluster richness parameter as defined in \citetalias{Remco2020}. The numerator is the sum of all cluster richness values, while the denominator is the sum of richness values for clusters with a stellar mass limit below the stellar mass of the galaxy the weight is being evaluated for.

Following \citet{Remco2013} and \citetalias{Remco2020}, our final sample consists of all galaxies with a stellar mass above the limit: $\log{\left[M_*/M_\odot\right]}> \max{\left(9.5,\log M_\mathrm{lim,\,cl}\right)}$, and with a $K_s$ magnitude larger than the limiting magnitude of their cluster (i.e. 1/Compl($K_s$) $<$ 1.25).  This results in 1,606 cluster members, with 833 designated as star-forming and 773 designated as quiescent. The number of star-forming and quiescent galaxies for each cluster are given in the final column of Table \ref{tab:Cluster_values}.

\section{Likelihood Function Formulation}\label{sec:LikelihoodChapter}
\subsection{Stellar mass function parameterisation}\label{sec:Extension}
We model the stellar mass functions (the galaxy number density per stellar mass interval) with a single \citet{Schechter1976} function of the form:
\begin{equation}\label{eq:schect}
    \phi(M)dM=\frac{dN}{dM}dM=
    \phi^*\left(\frac{M}{M^*}\right)^{\alpha}\exp\left(-\frac{M}{M^*}\right)\,\frac{dM}{M^*}.    
\end{equation}

\noindent To instead consider the number of galaxies per logarithmic mass interval, $\phi^\prime=dN/d\log{M}$, the form becomes:
\begin{equation}\label{eq:dlogM}
    \phi^\prime(M)\,d\log{M}=
    \ln{10}\,\phi^*\left(\frac{M}{M^*}\right)^{\alpha+1}\exp\left(-\frac{M}{M^*}\right)\,d\log{M}    
\end{equation}
\citep[e.g.][]{Annun2014,Remco2020} or, considering a logarithmic mass ($M^\prime=\log(M)$), the form becomes:
\begin{equation}\label{eq:logM}
    \phi^\prime(M^\prime)dM^\prime=
    \ln{10}\,\phi^*\left[10^{(M^\prime-M^{\prime*})(\alpha+1)}\right]\exp\left[-10^{(M^\prime-M^{\prime*})}\right]dM^\prime    
\end{equation}
as in \citet{Remco2018}.
While Equation \ref{eq:schect} is the form used in this work, all three forms are valid for use in the likelihood function shown in \S~\ref{sec:Mass_Likelihood}. \\

In our model, the parameters $M^*$ and $\alpha$, which parameterise the exponential cut-off and the low-mass slope, respectively, are extended to become functions of cluster-centric radius, $R$ and redshift, $z$, with a general form
\begin{align}
    \alpha&=\alpha(R,z) \\
    M^\ast&=M^*(R,z).
\end{align}
We propose a functional form for them as a linear combination of redshift and radius terms
\begin{align}
\alpha(r',z')&=\alpha_\circ +\alpha_R r^\prime+\alpha_z z^\prime+\alpha_{Rz} z^\prime r^\prime \label{alpha_eq} \\ 
    \log{M^\ast(r',z')}&=\log{M^\ast_\circ}+M^\ast_R r^\prime+M^\ast_z z^\prime +M^\ast_{Rz} z^\prime r^\prime, \label{M*_eq}
\end{align}
where $r^\prime=R/R_{200}-1$, and $z^\prime=(z-z_\circ)$. This ensures that the SMF parameters reduce to $M_\circ^\ast$ and $\alpha_\circ$ when $R=R_{200}$ and $z=z_{\circ}$. We centre our sample at $z_\circ=1.1$, the mean redshift of our clusters.

\subsection{Likelihood Function}\label{sec:Likelihood}

We define our model likelihood function as $\mathcal{L}=\prod_{i=1}^N{\mathcal{L}_i}$, where ${\mathcal{L}_i}$ is the likelihood calculated for each galaxy and is constructed as a product of two terms, which independently constrain the shape of the mass functions and the population normalisation for each galaxy type, respectively:
\begin{equation}\label{joint_likelihood}
    \mathcal{L}_{i,x}=\mathcal{L}_{M,i,x}\times\mathcal{L}_{R,i,x},
\end{equation}
where $i$ is the galaxy index, and the galaxy type (quiescent or star-forming) is specified by $x=\{Q,SF\}$.  We present each term separately below.

\subsubsection{Stellar mass dependence, $\mathcal{L}_{M,i,x}$}\label{sec:Mass_Likelihood}
The first term of Eq. \ref{joint_likelihood} considers a galaxy in a given cluster $cl$, with a redshift $z$, at a fixed radius $R$, and determines how likely that galaxy would be drawn from a mass distribution with a given set of SMF shape parameters. Following \citet{MK86} and \citet{OHE} we have the form
\begin{equation}\label{eq:Li}
    \mathcal{L}_{M,i}=\frac{\phi(M_i)dM}{\int_{M_{\rm lim}}^\infty{\phi(M)dM}}=\frac{\frac{1}{M^*}\left(\frac{M}{M^*}\right)^{\alpha}\exp\left(-\frac{M}{M^*}\right)dM}{\Gamma\left(1+\alpha,M_{\rm lim}/M^*\right)}.
\end{equation}
Note that the normalisation parameter, $\phi^*$ (in Eq. \ref{eq:schect}), cancels out, leaving a function that depends only on the parameters $M^\ast$ and $\alpha$.  $dM$ is a constant that is irrelevant for maximizing the likelihood, so we drop it for simplicity.

To reiterate, while not explicitly shown in the equation above, the parameters $\alpha$ and $M^*$ are functions of cluster-centric radius $R$, redshift $z$, and galaxy type $x$, as described in \S~\ref{sec:Extension}.

\subsubsection{Radial dependence, $\mathcal{L}_{R,i,x}$}\label{sec:Radius_Likelihood}
The second term of Eq. \ref{joint_likelihood} gives the likelihood of finding galaxy $i$, of type $x$, at a given radius $R$, in cluster $cl$.  This is assumed to depend only on the radial distribution of that galaxy type, in that cluster -- not the stellar mass. Like Eq. \ref{eq:Li}, we begin with the general form:
\begin{equation}
    \mathcal{L}_{R,i,x}=\frac{2 \pi R_i \; \varsigma(c_x,R_{i})dR}{\sum\limits_{\scriptscriptstyle k=Q,SF} \int_0^{R_{\rm max,cl}}2\pi R \; \varsigma(c_k,R) dR}.
\end{equation}

We assume the radial profile $\varsigma$, as a projected NFW \citep{LM2001}, with a different concentration parameter for the quiescent and star-forming populations: $c_Q$ and $c_{SF}$, respectively. To separate out the normalisation from the radial dependence we write the projected surface density as
\begin{equation}\label{eq:proj_expand}
\varsigma(c_x,R_i)=\Sigma^\ast_x\Sigma(c_x,R_i).
\end{equation}

Having a different profile for each population expands the denominator to be the sum of the integrals of these profiles, with the form

\begin{align}
    \mathcal{L}_{R,i,x}=&\frac{2\pi \Sigma^\ast_x \Sigma(c_x,R_i)R_idR}{\int_0^{R_{\rm max,cl}}2\pi \Sigma^\ast_{Q} \Sigma(c_{Q},R)R dR+\int_0^{R_{\rm max,cl}}2\pi \Sigma^\ast_{SF} \Sigma(c_{SF},R)R dR} \nonumber \\ =&f_q(cl_i)\frac{\Sigma^\ast_x }{\Sigma^\ast_{Q}}\frac{\Sigma(c_x,R_i)R_idR}{\int_0^{R_{\rm max,cl}}\Sigma(c_{Q},R)R dR},
\end{align}
where
\begin{align}\label{eq:quenched_frac}
f_q(cl_i)=&\frac{\int_0^{R_{\rm max,cl}}2\pi \Sigma^\ast_{Q} \Sigma(c_{Q},R)R dR}{\int_0^{R_{\rm max,cl}}2\pi \Sigma^\ast_{Q} \Sigma(c_{Q},R)R dR+\int_0^{R_{\rm max,cl}}2\pi \Sigma^\ast_{SF} \Sigma(c_{SF},R)R dR} \nonumber \\ =&
\frac{\int_0^{R_{\rm max,cl}} \Sigma(c_{Q},R)R dR}{\int_0^{R_{\rm max,cl}} \Sigma(c_{Q},R)R dR+\frac{\Sigma^\ast_{SF}}{\Sigma^\ast_{Q} } \int_0^{R_{\rm max,cl}} \Sigma(c_{SF},R)R dR},
\end{align}
defines the quenched fraction for cluster $cl_i$. The radial variation in the quenched fraction is then determined by the concentration parameters.

Writing out the likelihood function explicitly for the quiescent and star-forming populations gives us the forms
\begin{align}
    \mathcal{L}_{R,i,Q}&=f_q(cl_i)\frac{\Sigma(c_{Q},R_i)R_idR}{\int_0^{R_{\rm max,cl}}\Sigma(c_{Q},R)R dR}\label{eqn-Lik_R_r}\\
    \mathcal{L}_{R,i,SF}&=\left[1-f_q(cl_i)\right]\frac{\Sigma(c_{SF},R_i)R_idR}{\int_0^{R_{\rm max,cl}}\Sigma(c_{SF},R)R dR}.\label{eqn-Lik_R_b}
\end{align}
It can be seen that the radial likelihood term depends only\footnote{The overall normalisation of the quiescent population for each cluster, $\Sigma^\ast_{\rm Q}$, could also be fit as a set of free parameters. That was not done within this work, but in principle it would allow for a self-consistent measurement of the richness of each cluster.}
on the parameters $f_q$, $c_{Q}$, and $c_{SF}$. The relative normalisation term $\Sigma^\ast_{\rm SF}/\Sigma^\ast_{\rm Q}$ is not explicitly present in Eqs. \ref{eqn-Lik_R_r} and \ref{eqn-Lik_R_b} but is subsumed by $f_q$. 

We further parameterise and expand the quenched fraction term to be a function of redshift and cluster velocity dispersion (which we use as a proxy for dynamical mass), with the form
\begin{equation}\label{eq:fq}
    f_q(cl)=f_{q,\circ}(1+z^\prime)^{q_a}(1+q_b\sigma^\prime)
\end{equation}
where $\sigma^\prime=\left(\sigma_i/\sigma_\circ\right)^3$ and $z^\prime=(z_i-z_\circ)$, with $z_i$ and $\sigma_i$ being the respective redshift and velocity dispersion of the cluster $cl_i$ corresponding to galaxy $i$. Again we take both $z_\circ$ and $\sigma_\circ$ to be representative of the median of the sample at $z_\circ=1.1$ and $\sigma_\circ=500\mbox{km/s}$.

The function $\varsigma(c,R)$ and its corresponding integral over $R$ have analytic expressions. We use the form from \citet[although a similar form can be found in \citealt{Bartelmann1996}]{LM2001}; however, we use the slightly different notation of Eq. \ref{eq:proj_expand}, with a variable change of $r=R/R_{200}$, with
\begin{equation}\label{eq:proj_NFW}
    \Sigma(c,r)=\frac{1-{|c^2{r}^2-1|}^{-1/2}h(1/cr)}{\left(c^2{r}^2-1\right)},
\end{equation}
where $c=R_{200}/r_s$ is the concentration parameter, and
\begin{equation}
    h(x)=\begin{cases}
    \cos^{-1}(x) & \text{if $r>\frac{1}{c}$}\\
    \cosh^{-1}(x) & \text{if $r\leq\frac{1}{c}$}\\
    \end{cases}.
\end{equation}

The integral over $R$ can be solved analytically, again using the form from \citet{LM2001}:
\begin{equation}
    \int_0^{R}{\Sigma(R)RdR}=\frac{R_{200}^2}{c^2}\left[\frac{h\left(\frac{1}{cr}\right)}{\left|c^2r^2-1\right|^{1/2}}+\ln{\left(\frac{cr}{2}\right)}\right],
\end{equation}

which, when evaluating to the limit of $R=R_{200}$ simplifies to
\begin{equation}
    \int_0^{R_{200} }{\Sigma(R)RdR}=\frac{R_{200}^2}{c^2}\left[\frac{h\left(\frac{1}{c}\right)}{\left|c^2-1\right|^{1/2}}+\ln{\left(\frac{c}{2}\right)}\right].
\end{equation}

\subsubsection{Combined Likelihood}\label{sec:combined_like}
The final likelihood is the product of the mass and radial likelihood terms. The likelihood for a quiescent galaxy is shown below, given in a logged form to match the output used in our fitting routine: 
\begin{align}
    \log{\mathcal{L}_{i,Q}}=&\log{\left[\frac{1}{M^*}\left(\frac{M_i}{M^*}\right)^{\alpha}\exp\left(-\frac{M_i}{M^*}\right)dM\right]}+\log{\left[\Sigma(c_{Q},R_i)R_i\right]} \nonumber \\ &-\log{\Gamma\left(1+\alpha,\frac{M_{\rm lim(cl)}}{M^\ast}\right)} \nonumber \\
    &+\log{f_q(cl)}-\log{\left[\int_0^{R_{\rm max,cl}}\Sigma(c_{Q},R)R dR\right]},
\end{align}
where the top two lines contain quantities that have to be evaluated for every galaxy (again remembering that $\alpha$ and $M^*$ are evaluated for the specific $R_i$ and $z_i$ of each galaxy), and the bottom line contains quantities that have to be evaluated once for each cluster.

For the star-forming galaxies, the form is slightly altered:
\begin{align}
    \log{\mathcal{L}_{i,SF}}=&\log{\left[\frac{1}{M^*}\left(\frac{M_i}{M^*}\right)^{\alpha}\exp\left(-\frac{M_i}{M^*}\right)dM\right]}+\log{\left[\Sigma(c_{SF},R_i)R_i\right]} \nonumber \\ 
    &-\log{\Gamma\left(1+\alpha,\frac{M_{\rm lim(cl)}}{M^\ast}\right)} \nonumber \\
    &+\log{[1-f_q(cl)]}-\log{\left[\int_0^{R_{\rm max,cl}}\Sigma(c_{SF},R)R dR\right]}.
\end{align}

This likelihood structure results in nine free parameters to be fit for each population (quiescent and star-forming), in addition to the three parameters in $f_q$. This results in a total of 21 parameters for the whole model. The priors and posteriors for each parameter are shown in Table \ref{tab:params}. The high number of parameters were not expected to all be informative for the sample size we have. As discussed in \S~\ref{sec:BIC}, we explore a minimal set of these parameters and test the significance of additional parameters using the Bayesian Information Criterion (BIC).

\subsubsection{Weight incorporation}
In the likelihood described here, we assume that the sample is complete and unbiased over the cluster areas $A$, and for all stellar masses $M$ above the relevant stellar mass limit. This requires weighting the photometric sample for incompleteness as described in \S~\ref{sec:Weights}, and these weights are incorporated in the likelihood as in \citetalias{Remco2020}:
\begin{equation}
    \log{\mathcal{L}}=\sum_{i=1}^{N_g}{w_i\left[{\log{\mathcal{L}_{M,i,x}}+\log{\mathcal{L}_{R,i,x}}}\right]}.
\end{equation}

\subsection{Model Validation}
We tested the validity of the model by generating mock data corresponding to a specific set of parameters, and assessing the ability of the model to recover those parameters. 

First, galaxies from a range of cluster-centric radii were chosen by drawing from the probability density functions (PDFs) of projected NFW profiles, according to the concentration parameters.  We sample a set of discrete redshifts corresponding to the simulated clusters, which are chosen to match the redshifts of the clusters in our data. The radius and redshift of each galaxy then specifies $\alpha$ and $M^*$ according to our parameterisation, and a stellar mass is drawn from the corresponding Schechter function PDF. This gives mass, radius, and redshift values for each galaxy in the mock sample.

The entire set of 21 parameters was then fit to each sample, regardless of the size of the chosen set of mock sample parameters. This allowed us to validate that the model could recover all parameters correctly, even if they were set to null values. The true parameter values were consistently recovered within 1-$\sigma$ of the fit uncertainties. Multiple realizations of the same mock sample parameter sets were fit to confirm there was no bias within the parameter recovery.

\subsection{Fitting the data}\label{sec:emcee}
We fit the 21 parameter model to the sample of 1,606 galaxies described in \S~\ref{sec:Data} using the MCMC Ensemble sampler \texttt{emcee} \citep{emcee}. The input for the model is the set of galaxy stellar masses, cluster-centric radii, and their parent cluster redshifts. Other inputs include the stellar mass limit and virial radius of each cluster. The fit was completed using 50 walkers, and 3,000 iterations after the burn-in period of 1,000 iterations. The posterior values and their 1-$\sigma$ distributions are given in Table \ref{tab:params}, along with the prior ranges they were evaluated on.

The priors were set as uniform distributions with conservative ranges such that they were not dominating the parameter space the posteriors converged on, as can be seen in the complete corner plot in Figure \ref{fig:full_corner}. The ranges and initial values for the parameters not part of the radial/redshift $\alpha/M^*$ expansion were motivated by the fit values of the models in \citetalias{Remco2020} and \citet{Biviano2021}. In addition, to ensure that the overall quenched fraction remains within the physically motivated bounds of [0,1], we introduce a hyperprior of the upper limit of the normalisation term $f_{q,\circ}$, which enforces the inequality $f_{q,\circ}<1/\left[(1+z')^{q_a}(1+q_b\sigma')\right]$ (see Eq. \ref{eq:fq}).

\section{Results}\label{sec:Results}
We begin this section with \S~\ref{sec:Cluster_properties}, where we examine some key integrated properties of the clusters, and compare these with previous works. In \S~\ref{sec:SMF_fits} we focus on the main objective of this work, and present the results on how the shape of the SMFs depends on the cluster-centric radius and on the redshift of the cluster members.

\subsection{Radial Profiles and Quenched Fraction}\label{sec:Cluster_properties}
\begin{figure}
\centering
\includegraphics[width=1.0\linewidth]{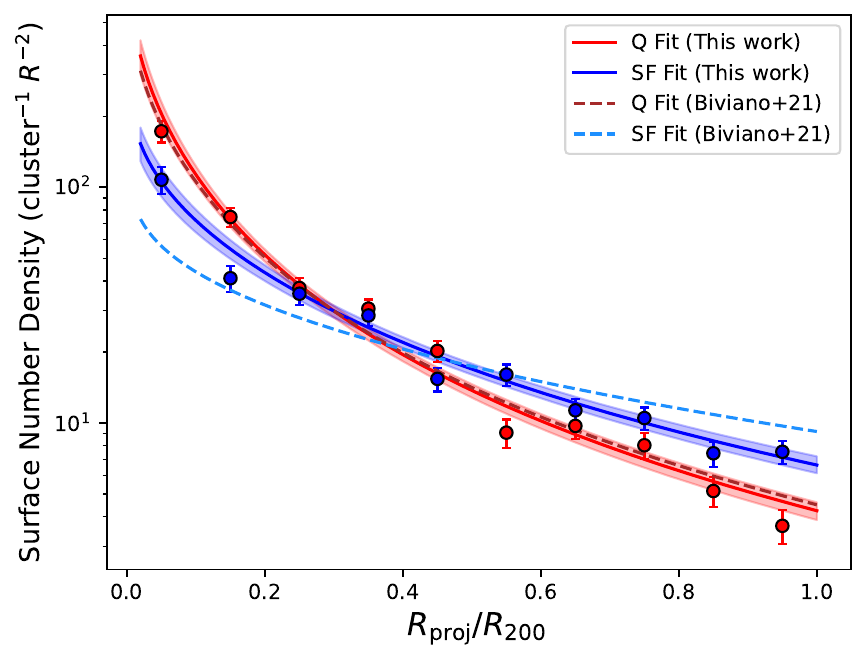}
\caption{The red and blue points show the surface number density of cluster galaxies in our sample as a function of cluster-centric radius for the quiescent and star-forming populations, respectively.  The shaded curves show the results of our full model and $1\sigma$ confidence region, determined by fitting the unbinned data.  These fits are compared with the profiles measured by \citet{Biviano2021} on a subset of this data, using a different method.}
\label{fig:surface_c}
\end{figure}

Figure \ref{fig:surface_c} shows the global surface density profiles of the quiescent and star-forming populations, defined by their respective concentration values. Overlaid are the profiles described by the best-fit concentration values in \citet{Biviano2021}, which was done on a subset of our cluster sample and with the same galaxy type definition. There is good agreement between their quiescent concentration value and ours (5.1 $\pm$ 0.9 and 6.0 $\pm$ 1.0, respectively), but a >2-$\sigma$ discrepancy between their star-forming value and ours (0.7 $\pm$ 0.4 and 2.07 $\pm$ 0.4, respectively). However, in their work they also split their data into two redshift bins, and for the higher redshift bin they fit a star-forming concentration value that is more consistent with ours (1.5 $\pm$ 1.0). Also, their sample does not include the three highest redshift GOGREEN clusters that our sample does (SpARCS-0219, SpARCS-1034, and SpARCS-1033). Finally, they use the BCG for the position of all their cluster centres, whereas we adopt different centres for SpARCS-1047 and SpARCS-1051. Nevertheless, the quiescent profile has a steeper slope than the star-forming in both sets of parameters, which corresponds to a quiescent-dominated core region ($f_q>0.5$), as shown explicitly in later figures.

\begin{figure}
\centering
\includegraphics[width=1.0\linewidth]{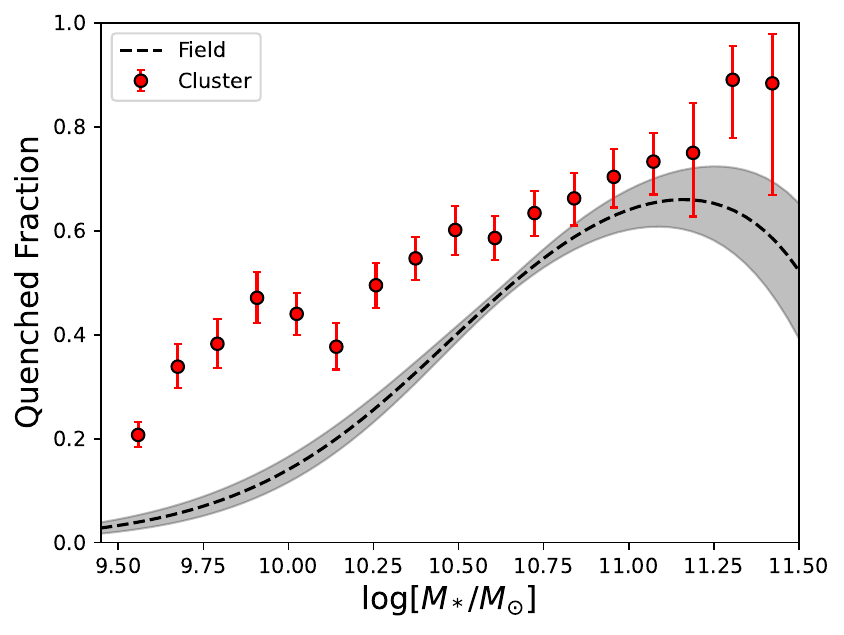}
\caption{Global quenched fraction dependence on stellar mass, for all radii $R<R_{200}$ and redshifts. The uncertainties on the binned points correspond to the binomial confidence intervals. The dashed line and grey shaded regions show the quenched fraction values of field galaxies in the redshift range of the sample, following the fits from \citet{McLeod2021}, and their 1-$\sigma$ uncertainties. The figure is omitting the three highest mass bins of the sample as neither the field nor the cluster quenched fractions are usefully constrained at those  masses.}
\label{fig:Fq_M}
\end{figure}

Figure \ref{fig:Fq_M} shows the quenched fraction within $R_{200}$ of the sample as a function of stellar mass. There is a strong correlation such that the quenched fraction increases from $\sim$20 per cent at the mass limit of the sample, to almost entirely quenched at the highest masses. This trend is consistent with the form found in \citetalias{Remco2020} and other studies \citep{Peng2010,Muzzin2012,Remco2018}. The values are compared to the quenched fractions from the field SMFs from \citet{McLeod2021}, evaluated at the mean redshift of our sample. The cluster quenched fraction is elevated above the field levels at all stellar masses, and is significantly elevated at $\log M^*<10.75$.

\begin{figure}
\centering
\includegraphics[width=1.0\linewidth]{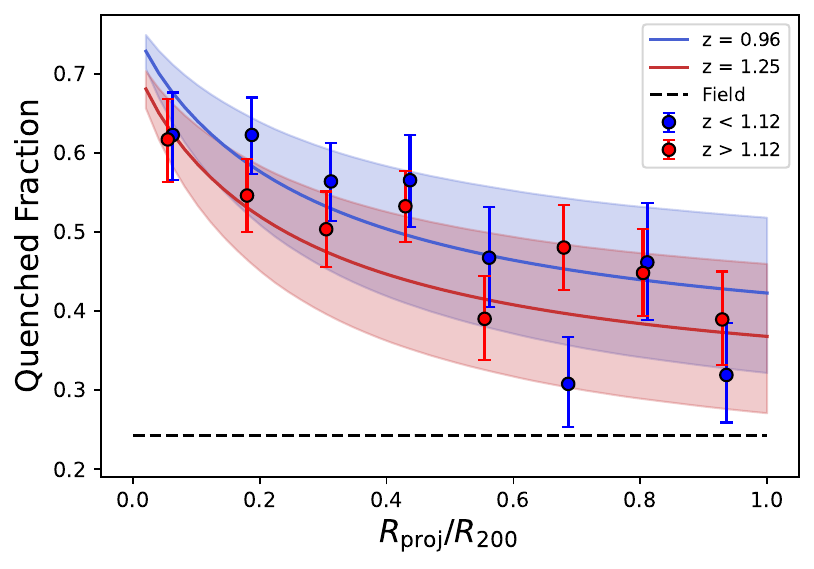}
\caption{Quenched fraction as a function of cluster-centric radius. The sample is split into two bins based on the mean redshift of the sample ($z\simeq1.12$). For clarity, the high redshift points are slightly horizontally offset. The quenched fraction profiles (Eq. \ref{eq:quenched_frac}) are evaluated at the mean redshift values their redshift bins. The uncertainties on the binned points are their corresponding binomial confidence intervals, and the shaded region around the fits are the 1-$\sigma$ parameter uncertainties. The field value is set to the global field quenched fraction value from \citet{McLeod2021}, at the mean redshift of our sample, and down to our stellar mass limit.}
\label{fig:Fq_R}
\end{figure}

Figure \ref{fig:Fq_R} shows the quenched fraction as a function of cluster-centric radius. The galaxies are split into two redshift bins, with the analytic function being evaluated for the mean redshifts of each. Both redshift bins show an inverse radial correlation, with the quenched fraction increasing with proximity to the cluster core. This is consistent with trends observed in this \citep[e.g.][]{Baxter2022} and other samples, including at $z=0$ \citep{Vulcani2013,Haines2015,Kawinwanichakij2017,Pintos-Castro2019}, and is qualitatively consistent with the observed trend for quenched fraction to increase with increasing local density \citep{Kauffmann2004,Peng2010}. The observed trend is compared with the global quenched fraction of the field (evaluated at the mean redshift and over the same stellar mass range as our sample),  from \citet{McLeod2021}. The quenched fraction of the outer regions of the clusters is $\sim 40\pm10$ per cent, larger than the field ($\sim 25$ per cent) but at low significance.  This difference increases rapidly towards the core, where the cluster quenched fraction reaches $\sim 70$ per cent.  The redshift split of the fit functions indicates a small ($<5$ per cent) but statistically insignificant normalisation increase in quenched fraction for all radii with a decrease in redshift. Further analysis of trends with redshift and cluster velocity dispersion are explored in Appendix~\ref{app-B}.

\begin{table}
\centering
\begin{tabular}{@{}lll@{}}
\toprule
\textbf{Parameter}\hspace{0.5cm} & \textbf{Prior}\hspace{1.0cm} & \textbf{Posterior} \\
\midrule
$\alpha_{\circ,Q}$ & [-10 , 10] & $-0.65_{-0.26}^{+0.29}$\\
$\alpha_{\circ,SF}$ & [-10 , 10] & $-1.23_{-0.20}^{+0.20}$\\
$\alpha_{R,Q}$ & [-10 , 10] & $-0.33_{-0.40}^{+0.42}$\\
$\alpha_{R,SF}$ & [-10 , 10] & $-0.05_{-0.34}^{+0.33}$\\
$\alpha_{z,Q}$ & [-10 , 10] & $1.54_{-1.52}^{+1.55}$\\
$\alpha_{z,SF}$ & [-10 , 10] & $0.40_{-1.14}^{+1.26}$\\
$\alpha_{Rz,Q}$ & [-10 , 10] & $1.20_{-2.34}^{+2.23}$\\
$\alpha_{Rz,SF}$ & [-10 , 10] & $0.22_{-1.99}^{+2.06}$\\
$M^*_{\circ,Q}$ & [5 , 30] & $10.66_{-0.12}^{+0.12}$\\
$M^*_{\circ,SF}$ & [5 , 30] & $10.79_{-0.15}^{+0.18}$\\
$M^*_{R,Q}$ & [-10 , 10] & $-0.26_{-0.18}^{+0.18}$\\
$M^*_{R,SF}$ & [-10 , 10] & $-0.24_{-0.30}^{+0.30}$\\
$M^*_{z,Q}$ & [-10 , 10] & $-0.52_{-0.61}^{+0.61}$\\
$M^*_{z,SF}$ & [-10 , 10] & $-0.26_{-0.85}^{+0.84}$\\
$M^*_{Rz,Q}$ & [-10 , 10] & $-0.12_{-0.90}^{+0.95}$\\
$M^*_{Rz,SF}$ & [-10 , 10] & $0.42_{-1.64}^{+1.58}$\\
$c_{Q}$ & [0 , 20] & $5.98_{-0.91}^{+1.09}$ \\
$c_{SF}$ & [0 , 20] & $2.07_{-0.39}^{+0.42}$ \\
$f_{q,\circ}$ & [0 , 1$^{(a)}$] & $0.49_{-0.03}^{+0.03}$\\
$q_a$ & [-5 , 5] & $-0.41_{-0.23}^{+0.25}$\\
$q_b$ & [-5 , 5] & $0.00_{-0.01}^{+0.01}$\\
\bottomrule
\end{tabular}
\caption{\label{tab:params}An overview of the 21 parameters, their prior ranges, and their best-fit posterior values with their 1-$\sigma$/68 per cent confidence limits. The prior ranges all correspond to flat priors. $^{(a)}$The upper range of the prior for $f_{q,\circ}$ is set as a hyperprior so the overall $f_q$ term is less than 1 (see \S~\ref{sec:emcee}).}
\end{table}

\subsection{Stellar Mass Function Fits}\label{sec:SMF_fits}
To demonstrate how well our model is able to fit the stellar mass functions of cluster populations as a function of cluster-centric radius and redshift, we first divide each population (quiescent and star-forming) into six sub-populations: two radial regimes ($0<R/R_{200}\leq0.45$ and $0.45<R/R_{200}\leq1.0$), and three redshift regimes ($0.86<z\leq1.05$, $1.05<z\leq1.25$, and $1.25<z\leq1.46$), all of which have approximately equal sample sizes. We plot the binned number density as a function of mass for each subset, and overlay the mass function from the full model but with the global $\alpha$ and $M^*$ parameters evaluated at the mean radius and redshift of each sub-sample. We also fit a Schechter function with no radial or redshift dependency (see Eq. \ref{eq:schect}) to each subset, which we call the `simple model', giving a comparison fit. Within these plots, the errors on the SMF binned points are Poisson errors with the relative magnitude of their weights taken into account.

\subsubsection{Star-Forming Population}
\begin{figure*}
\centering
\includegraphics[width=0.95\linewidth]{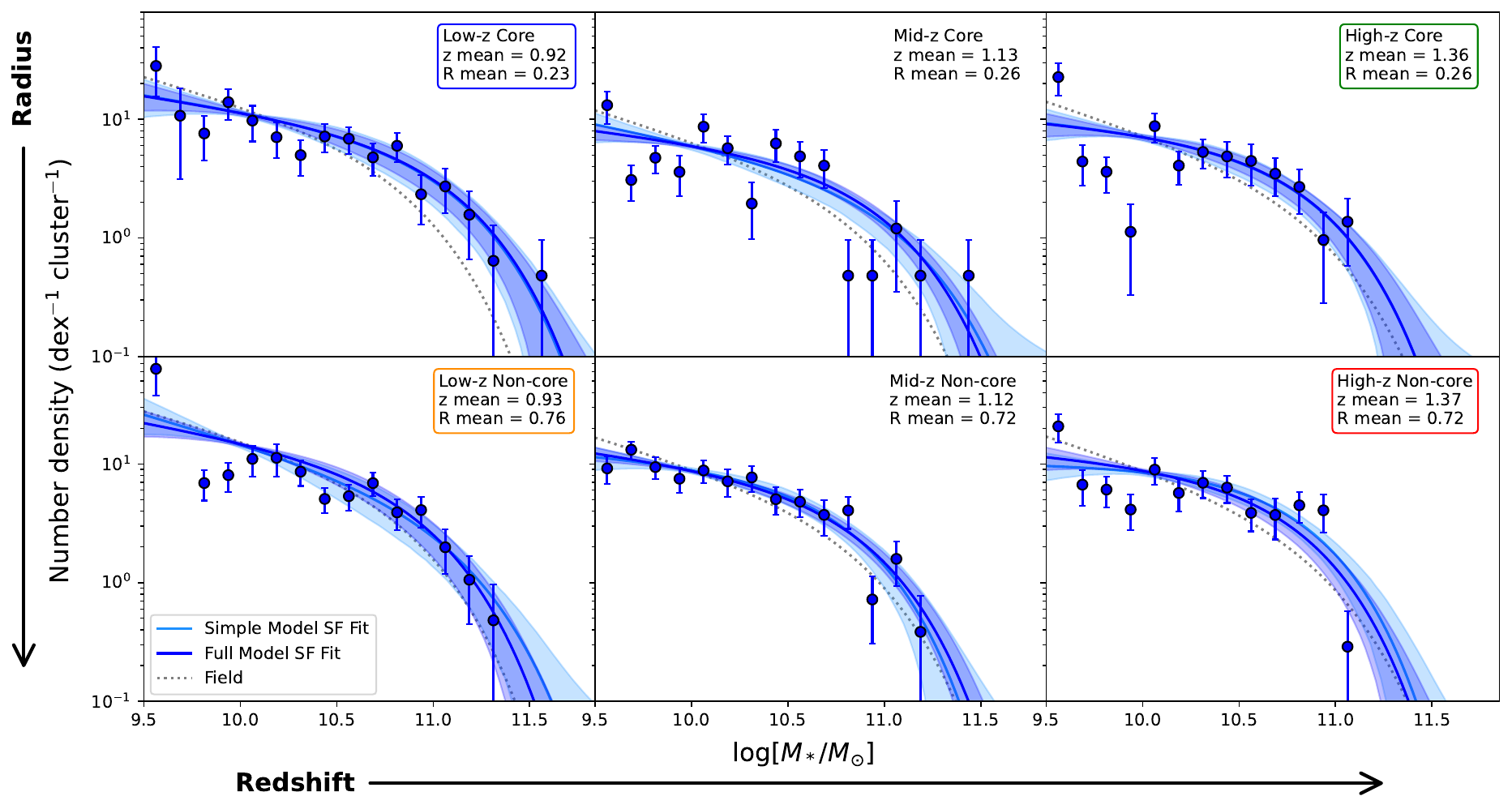}
\caption{Binned number density of star-forming galaxies against their stellar mass for two radial bins and three redshift bins. Overlaid is the comparison of the complete model (the dark-blue `Full Model' fit) evaluated at the mean $z$ and $R/R_{200}$ of each group (which are stated in each sub-figure). For comparison, we show a non-radial and non-redshift dependent Schechter function fit for just the population in each subgroup (the light-blue `Simple Model' fit). The colours of the title boxes match the colours used for the respective populations in Figure \ref{fig:SF_params}. The two models are fully consistent with one another, but the full model is better constrained, as expected since it is derived from the full dataset. Also shown are the SF field SMFs from \citet{McLeod2021}, evaluated at the mean redshift of, and normalised to, each subgroup, in order to have a visual comparison between their shapes.}
\label{fig:Blue_SMFs}
\end{figure*}

Figure \ref{fig:Blue_SMFs} shows the six sub-populations of the star-forming galaxies, with binned galaxy number density by stellar mass. Overlaid are both the full model and simple model fits to each sub-sample. The two fits are consistent with one another, but the full model fits are better constrained, especially at the low and high mass ends. This is a reflection of the additional information the complete model has from its access to the entire sample.

\begin{figure}
\centering
\includegraphics[width=1.0\linewidth]{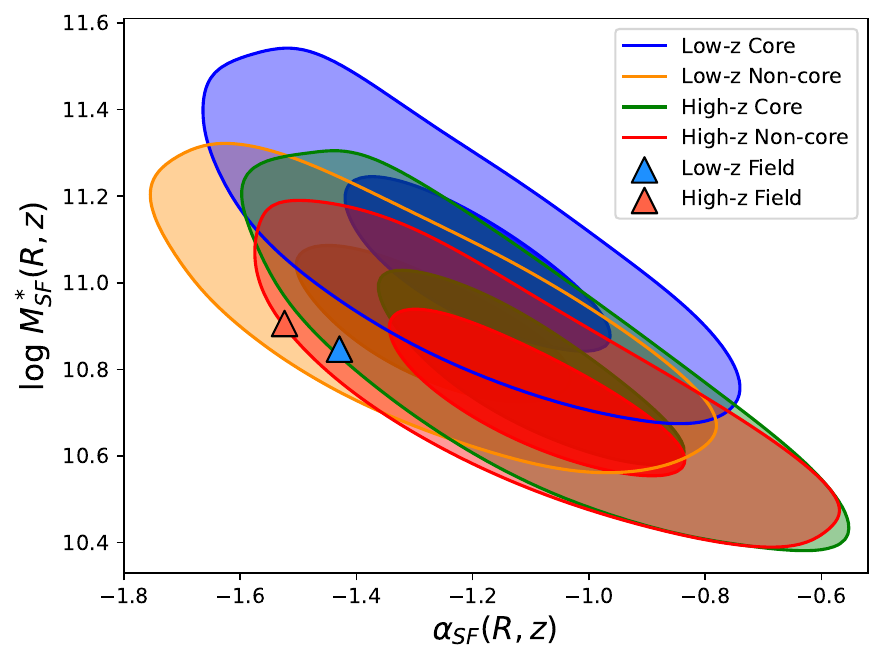}
\caption{The best-fit $\alpha$ and $M^*$ values and their 1-$\sigma$ and 2-$\sigma$ uncertainties for the star-forming population. The parameter values are evaluated for the mean radial and redshift values of the four edge sub-populations in Figure \ref{fig:Blue_SMFs}, to show the extent of the parameter space explored. The star-forming field values from \citet{McLeod2021} are also shown, with the light blue and pink triangles corresponding to the lowest and highest redshifts of our sample redshift range, respectively. The field 1-$\sigma$ uncertainties are the same scale as the plot icons. There is no significant variation in the shape of the SMF for this population, as a function of environment or redshift.}
\label{fig:SF_params}
\end{figure}

Figure \ref{fig:SF_params} shows the $\alpha$ and $M^*$ best-fit values and their 1-$\sigma$ and 2-$\sigma$ contours. The parameters are shown evaluated at the mean $z$ and $R/R_{200}$ for each of the four edge populations in Figure \ref{fig:Blue_SMFs}, with \emph{Low-z} corresponding to the left-most column, \emph{High-z} to the right-most column, and \emph{Core} and \emph{Non-core} corresponding to the top and bottom rows, respectively. These groups were chosen to show the extent of the explored parameter space of the full model. Also shown are the values of the field (evaluated at the edge of the sample redshift range using the \citet{McLeod2021} fits), with the statistical uncertainties of the field given in their error bars (note that for the star-forming population the errors are on the scale as the plot icons). These uncertainties do not include systematic uncertainties that might be present, for example in the definitions of star-forming and quiescent from \emph{UVJ} colours). For example, results from \citet{Muzzin2013}, who use a different \emph{UVJ} selection cut, find the SF $\alpha$ field values to be -1.2, compared to $\sim$-1.5 from \citet{McLeod2021}. Considering this, the figure shows that the cluster $\alpha$ and $M^*$ values are not only consistent with the field values, but are also consistent through both the redshift and radial ranges. These results indicate that the star-forming populations in the clusters do not significantly depend on environment or redshift within the sample, and are consistent with their field counterparts.

\subsubsection{Quiescent Population}
\begin{figure*}
\centering
\includegraphics[width=0.95\linewidth]{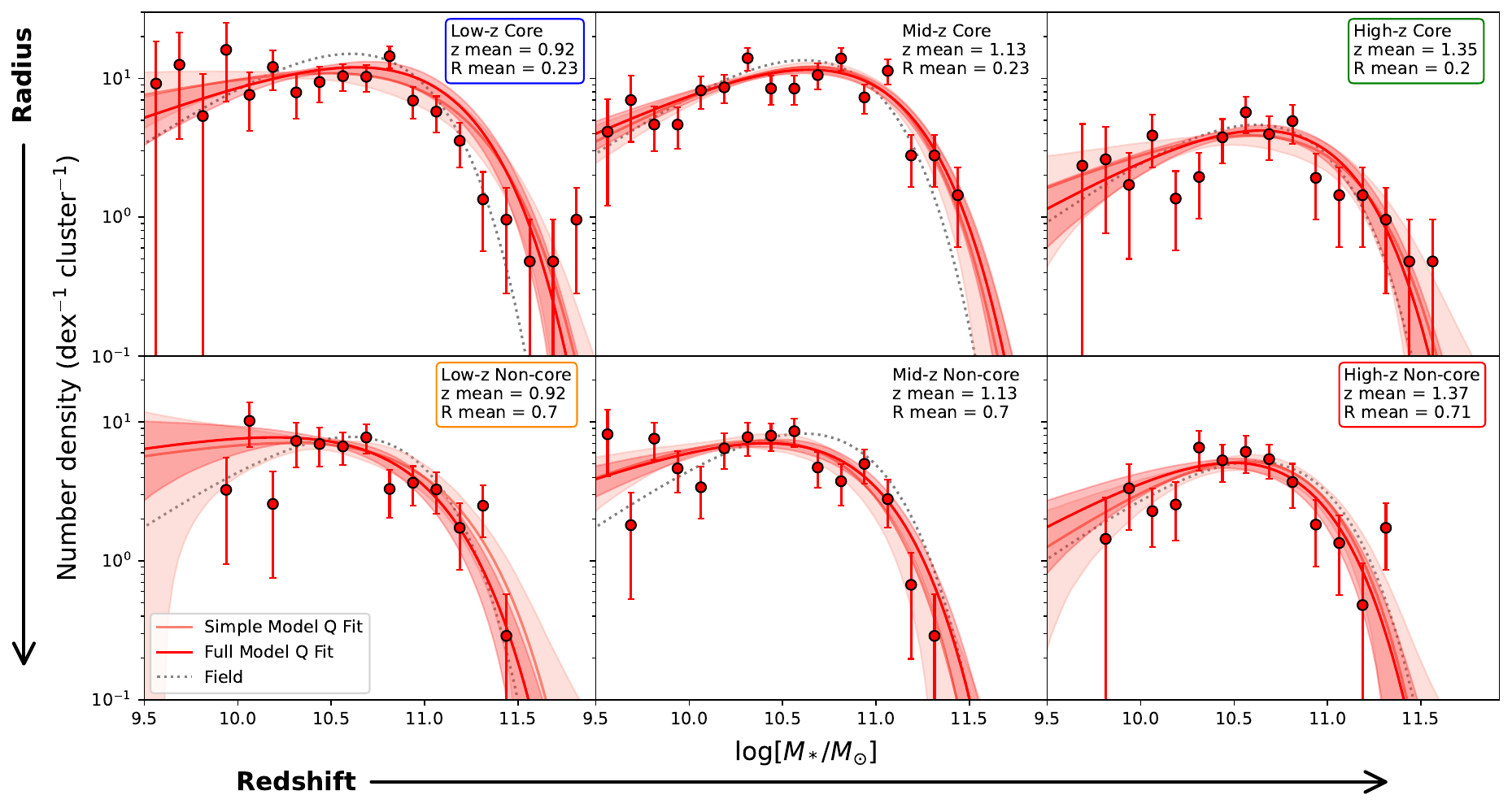}
\caption{Same form of Figure \ref{fig:Blue_SMFs}, but showing the binned number density against stellar mass for the quiescent galaxy population (again through two radial bins and three redshift bins). Overlaid is the comparison of the complete model (the red `Full Model' fit) evaluated at the mean $z$ and $R/R_{200}$ of each group (which are stated in each sub-figure). For comparison, we show a non-radial and non-redshift dependent Schechter function fit for just the population in each subgroup (the pink `Simple Model' fit). The colours of the title boxes match the colours used for the respective populations in Figure \ref{fig:Q_params}. There is good agreement between the different model fits and the binned data. Also shown are the Q field SMFs from \citet{McLeod2021}, evaluated for the mean redshift of, and normalised to, each subgroup, in order to have a visual comparison between their shapes.}
\label{fig:Red_SMFs}
\end{figure*}

Figure \ref{fig:Red_SMFs} follows the same form as Figure \ref{fig:Blue_SMFs}, but instead shows the six sub-populations of the quiescent galaxies, along with the full model and simple model fits for each sub-sample. Again both fits are consistent with each other, with the major difference that the full model fits are better constrained than the individual simple model fits, in particular at the low mass end. This is due to the additional information that the model gained from its access to the entire sample.

\begin{figure} \vspace{0.2cm} 
\centering
\includegraphics[width=1.0\linewidth]{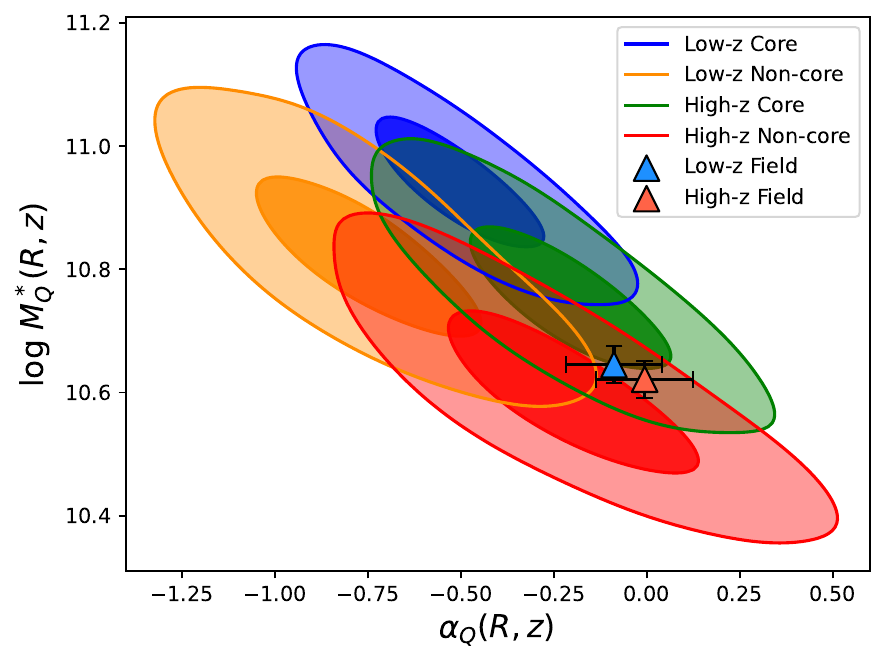}
\caption{Same form of Figure \ref{fig:SF_params}, but showing the best-fit $\alpha$ and $M^*$ values for the quiescent population, along with their 1-$\sigma$ and 2-$\sigma$ uncertainties. The parameter values are evaluated for the mean radial and redshift values of the four edge sub-populations in Figure \ref{fig:Red_SMFs} to show the extent of the parameter space explored. The quiescent field values from \citet{McLeod2021} (and their 1-$\sigma$ uncertainties) are also shown, with the light-blue and pink triangles corresponding to the lowest and highest redshifts of our sample redshift range, respectively. There are no significant variations in model parameters for this full model, though there is a weak trend for a dependence on cluster-centric radius (core vs non-core).}
\label{fig:Q_params}
\end{figure}

In the same form as Figure \ref{fig:SF_params}, Figure \ref{fig:Q_params} shows the overall $\alpha$ and $M^*$ best-fit values for the individual Schechter parameters for the quiescent population. The values and their 1-$\sigma$ and 2$\sigma$ contours are again subdivided into four groups, evaluated for the mean $z$ and $R/R_{200}$ of the edge sub-populations in Figure \ref{fig:Red_SMFs}. The subpopulations are consistent with the \citet{McLeod2021} field values (again taking into account the systematic errors - although there is more consistency in the literature for these values, with \citet{Muzzin2013} finding a quiescent $\alpha$ values of -0.17), however, there is $>1\sigma$ evidence for a trend with radius.  At all redshifts, quiescent galaxies in the core exhibit a slightly more positive $\alpha$ and a larger $M^\ast$ than cluster galaxies outside the core. This will be explored further in the next section.

\subsection{Constraining Parameters and $\Delta$BIC}\label{sec:BIC}
As mentioned in \S~\ref{sec:combined_like} and as is evident through the uncertainties on some of the parameters in Table \ref{tab:params}, not all of the parameters we fit for are informative. In particular, many of the parameters that describe the radial and redshift dependence of $\alpha$ and $M^*$ are consistent with zero. Since the inclusion of extra parameters lead to increased uncertainties, this may obscure any physically relevant trends. 

Thus, in order to constrain the parameter set to the most significant parameters of the model, we first limit the model to a simplified base set of parameters that are expected to be informative: $\left[c_r,c_b,\alpha_{\circ,r},\alpha_{\circ,b},M^*_{\circ,r},M^*_{\circ,b},f_{q,\circ}\right]$. This base model assumes a global $\alpha$ and $M^*$ value that is independent of radius and redshift in both the star-forming and quiescent populations, a concentration value for the radial profiles of each population, and a fixed quenched fraction term that is independent of redshift or velocity dispersion. Starting from this base model, we then add additional parameters and validate their significance through the difference in the BIC measurements of each model ($\Delta$BIC). The BIC quantifies \emph{significant} improvement of the model through the addition of parameters by penalizing models by the size of their parameter space, with a form of $k*\ln(N) - 2\ln(\mathcal{L})$, where $N$ is the sample size, $k$ is the number of parameters, and $\mathcal{L}$ is the likelihood of that specific model \citep{Schwarz1978}.

A variety of parameter combinations were explored, with their BIC and $\Delta$BIC values given in Table \ref{tab:BIC}. A rough interpretation of $\Delta$BIC values is that negative values indicate no evidence for significant improvement of the model with additional parameters, 0$<$$\Delta$BIC$<$2 indicates weak evidence of significant improvement, 2$<$$\Delta$BIC$<$6 indicates moderate evidence of significant improvement, 6$<$$\Delta$BIC$<$10 indicates strong evidence of significant improvement, and $\Delta$BIC$>$10 indicates very strong evidence of significant improvement with additional parameters \citep{Raftery1995}. 

We find only two parameter configurations that yield a positive $\Delta$BIC: the addition of the radial component in the quiescent $\alpha$ expansion, and the addition of the radial component in the quiescent $M^*$ expansion, which both individually present moderate evidence for a significant improvement with their addition. We interpret this as moderate evidence that the shape of the quiescent SMF varies with location in the cluster, as already indicated in Figure~\ref{fig:Q_params}. Because of the strong degeneracy between the $\alpha$ and $M^*$ parameters, it is difficult to isolate which one drives this trend. In fact, including both parameters in the model results in a slightly negative $\Delta$BIC relative to the Base model. However, despite their degeneracy, the parameters have different physical interpretations and only adding one of them could lead to misleading conclusions. We therefore elect to explore a model in which both parameters are free.

\begin{figure}
\centering
\includegraphics[width=1.0\linewidth]{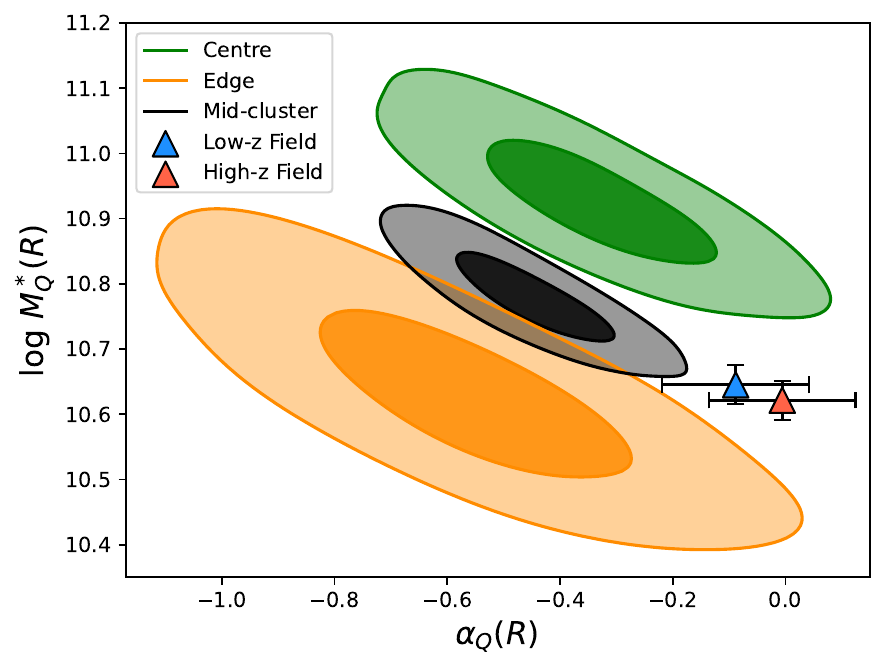}
\caption{Same form as Figure \ref{fig:Q_params}, but showing the best-fit $\alpha$ and $M^*$ values for the quiescent population (and their 1-$\sigma$ and 2-$\sigma$ uncertainties) for the model with only radial parameters for the quiescent SMF shapes (the `Base + $\alpha_{R,r}$ + $M^*_{R,r}$' model within Table \ref{tab:BIC}). The centre, mid-cluster, and edge populations are evaluated at 0, 0.5, and 1 $R_{200}$ of the clusters. The quiescent field values from \citet{McLeod2021} (and their 1-$\sigma$ uncertainties) are shown, with the light-blue and pink triangles corresponding to the lowest and highest redshifts of our sample redshift range, respectively. In this restricted model, a strong dependence with environment is observed, with $>2\sigma$ significance across the virialised range.}
\label{fig:Q_radial_model}
\end{figure}

Figure \ref{fig:Q_radial_model} shows the quiescent $\alpha$ and $M^*$ values for the model that only has the radial Schechter expansion parameters (the `Base + $\alpha_{R,r}$ + $M^*_{R,r}$' model in Table \ref{tab:BIC}). Here the \emph{Centre} and \emph{Edge} regions are evaluated at the extreme radial limits ($R/R_{200}=0$ and $R/R_{200}=1$), in order to show the maximum extent of the explored parameter space (we define the radial extent of the cluster core as $R<0.3R_{200}$). We now observe a $2.1\sigma$ separation between these two populations. We also show the parameter values at half the virial radius, corresponding to a \emph{Mid-cluster} population. This demonstrates both the smooth transition of the trends, and how the parameter error constraints change across our sample range. Specifically, the parameters are better constrained for the middle population, because this is where most of the data is. Importantly, the parameters for the Centre population are inconsistent with the field at the $3.8\sigma$ and the $3.6\sigma$ levels for the \emph{low-z} and \emph{high-z} values, respectively.

\begin{table}
\centering
\begin{tabular}{@{}lll@{}}
\toprule
\textbf{Parameter Combination} & \textbf{BIC Value} & \textbf{$\Delta$BIC} \\
\midrule
Base & 5096.20 & N/A \\
Base + $\alpha_{R,Q}$ & 5092.93 & \textbf{3.27} \\
Base + $M^*_{R,Q}$ & 5090.23 & \textbf{5.97} \\
Base + $\alpha_{z,Q}$ & 5103.52 & -7.32 \\
Base + $M^*_{z,Q}$ & 5101.86 & -5.66 \\
Base + $\alpha_{Rz,Q}$ & 5103.48 & -7.28 \\
Base + $M^*_{Rz,Q}$ & 5102.51 & -6.31 \\
Base + $\alpha_{R,SF}$ & 5102.52 & -6.32 \\
Base + $M^*_{R,SF}$ & 5101.67 & -5.47 \\
Base + $\alpha_{z,SF}$ & 5103.40 & -7.20 \\
Base + $M^*_{z,SF}$ & 5102.40 & -6.20 \\
Base + $\alpha_{Rz,SF}$ & 5103.25 & -7.05 \\
Base + $M^*_{Rz,SF}$ & 5102.39 & -6.19 \\
Base + $q_a$ & 5098.34 & -2.14 \\
Base + $q_b$ & 5101.56 & -5.36 \\
Base + $\alpha_{R,Q}$ +  $M^*_{R,Q}$ & 5097.12 & -0.92 \\
Base + $\alpha_{z,Q}$ +  $M^*_{z,Q}$ & 5108.52 & -12.32 \\
All parameters & 5178.85 & -82.65 \\
\bottomrule
\end{tabular}
\caption{\label{tab:BIC}Select combinations of model parameter subsets, along with their BIC value and their $\Delta$BIC value relative to the `Base' combination (given in \S~\ref{sec:BIC}).}
\end{table}

\subsubsection{Total stellar mass functions}
\begin{figure}
\centering
\includegraphics[width=1.0\linewidth]{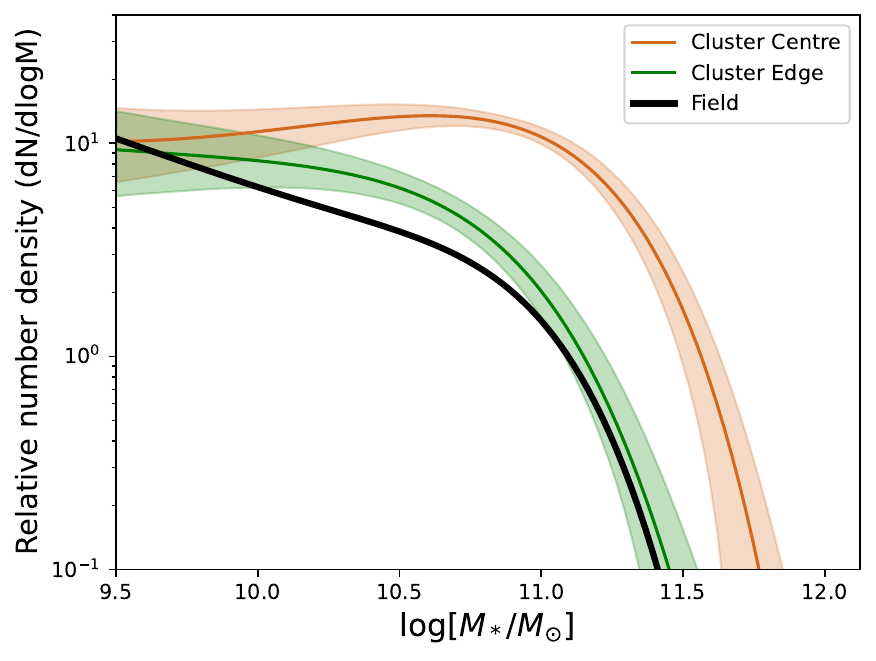}
\caption{The total SMFs of the cluster centre (brown) and the cluster edge (green) with their respective 1-$\sigma$ uncertainties. Also shown in black is the total field SMF based on the fits from \citet{McLeod2021}. All three SMFs are normalised to have the same integrated SF SMF component. The curves therefore reflect the different shapes in total SMF, and also the different overall quenched fractions. At the cluster edge, the SMF is similar to the field SMF, while the cluster centre SMF has a significantly different shape from both.}
\label{fig:Total_SMF}
\end{figure}

The results from Figure \ref{fig:Fq_R} show an increased quenched fraction compared to the field throughout the clusters, and especially in the core. This, coupled with the similarities of the individual Schechter parameters of the quiescent SMFs with the field means the total SMFs of the clusters must also vary with environment. This is shown in Figure \ref{fig:Total_SMF}, where we present the combined star-forming and quiescent SMFs of the centre and edge populations, and the field. This figure shows that the edge SMF is similar in form to that of the field, while the centre has a significantly different shape, with a much greater fraction of massive galaxies. This suggests the centre population can not simply be formed from a direct quenching of the field population, something we explore further in \S~\ref{sec:Toy_Models}.

\section{Discussion}\label{sec:Discussion}
\subsection{Comparison to Literature}\label{sec:Literature}
Through Figures \ref{fig:Fq_M}, \ref{fig:Fq_R} (and \ref{fig:Fq_z}), we find that the quenched fraction in these massive clusters is dependent on a galaxy's stellar mass, cluster-centric radius, and redshift. This is consistent and expected based on previous findings \citep{Dressler1980,Peng2010,Tomczak2014,Tomczak2017,Haines2015,Nantais2016,Jian2017,Guglielmo2019,Pintos-Castro2019,Roberts2019,Stott2022}, including previous analysis of this same sample \citep{Muzzin2012,Muzzin2013,Remco2020,Baxter2022,Baxter2023}. Using a simple model where mass-quenching and environmental-quenching are separable \citep[e.g.][]{Peng2010,Muzzin2012}, it is assumed that the excess of quenched galaxies in clusters is built up through a mass independent conversion of the star-forming population. If this model is correct, it is necessary that the quiescent SMFs of cluster galaxies should be different from the field mass functions. Observations of low redshift clusters indeed find a SMF difference, and this separable model is able to reproduce the observations in the local Universe \citep{Peng2010} (but see \citet{Vulcani2013} for alternative results).

The initial analysis of GOGREEN clusters at higher redshift presented a different result. \citetalias{Remco2020} showed that both the quiescent and star-forming cluster SMFs at redshifts $1.0<z<1.4$ have no measurable difference from the shapes of their respective field populations. The conclusion they drew was that there must be some shared mechanism in both environments that results in the observed mass-dependent quenching, but at an enhanced rate in clusters to achieve the elevated quenched fraction. They explore possibilities including an `early mass-quenching' model with earlier formation times of the cluster member progenitors. 

In principle, evidence for this early mass quenching could be found by measuring the ages of quiescent galaxies. This measurement was made by \citet{Webb2020}, who did confirm that cluster quiescent galaxies are somewhat older than those in the field, as expected in this model, and different from what is expected in an infall-driven environmental quenching scenario. However, the sample was too small to measure this difference as a function of cluster-centric radius. 

In contrast, there is also evidence that some quenching must occur later, during the assembly of the cluster. \citet{Werner2022} show that massive galaxies in GOGREEN are preferentially being quenched in the infall regions around the clusters. Also, \citet{Edward2024} analysed protoclusters at $z>2$ and found that a substantial amount of quenching must occur between then and $z\sim 1.5$ to match the quiescent stellar mass functions in GOGREEN clusters. Finally, \citet{Baxter2022,Baxter2023} also analysed these same clusters and showed the strong radial and mass dependence of the quenched fraction could be modelled with a mass-dependent, environmental quenching mechanism. In particular, \citet{Baxter2023} found two solutions, corresponding to a `core-quenching' scenario and a `starvation' non-core scenario, though they ruled out the former as a dominant mechanism using the phase-space distribution of transition galaxies. However, they did not explore in detail the extent to which this could explain the shape of the quiescent SMF as a function of environment. It is evident from these results that multiple quenching modes may have built up the quenched population in these clusters, and the ability of our model to fit trends through radius and redshift is well suited to distinguish between them.

In Figure \ref{fig:SF_params} we show that the fits for $\alpha_\mathrm{SF}$ and $M^*_\mathrm{SF}$ are invariant as a function of radius and redshift, as well as consistent with their related field values. This result is consistent with expectations from both the previously mentioned quenching models, and has been seen in previous work through various redshift and environmental conditions \citep{Calvi2013,Ilbert2013,Vulcani2013,Annun2014,Bundy2017,Papovich2018,Kawinwanichakij2020}.

However, when we allow $\alpha_Q$ and $M^\ast_Q$ to vary with cluster-centric radius, we do find a moderately significant difference with environment, as shown in Figure \ref{fig:Q_radial_model}. A significant difference in SMF shape relative to the field is observed, but only at the cluster centre.  This difference was therefore not apparent in the global SMFs analysed by \citetalias{Remco2020}. It is also relevant that when we allow the extra freedom for the model to account for an important radial dependence, the parameter uncertainties increase. The restricted (radial-independent) model of \citetalias{Remco2020} would therefore also lead to underestimated uncertainties and an overestimate of the significance of their null result. Finally, we note that recent work \citep{Gully2025}, which took advantage of deeper, narrow-band data on a subset of four GCLASS clusters, did find evidence for a significant change in the low-mass slope of the quiescent SMF relative to the field. 

While the redshift dependence of the full model is not found to be significant, it is still important to understand how our results fit in the context of the global quiescent population. \citet{Papovich2018} analysed a sample of galaxies in various environments over a  redshift range of $0<z<2$. They use survey data from ZFOURGE \citep{Straatman2016} and NMBS \citep{Whitaker2011}, which both cover random fields over a combined area of $\sim$1,800 arcmin$^2$. They also include results from \citet{Tomczak2014}, which independently measured the SMF parameters of the total quiescent population using an earlier version of the ZFOURGE dataset. The results from their analysis show that the shape of quiescent SMFs in low-density environments do not evolve with redshift, while the SMF shape in higher densities begins to deviate from the `field' at $z<1$ (similar results are also found in \citet{Davidzon2016}). Our study probes the redshift regime where this transition occurs. It should be noted however, that since their data cover field galaxies and not rich structures as our sample does, their highest density quartile is more akin to our lowest density regime (the cluster non-core/edge).

\begin{figure*}
\centering
\includegraphics[width=1.0\linewidth]{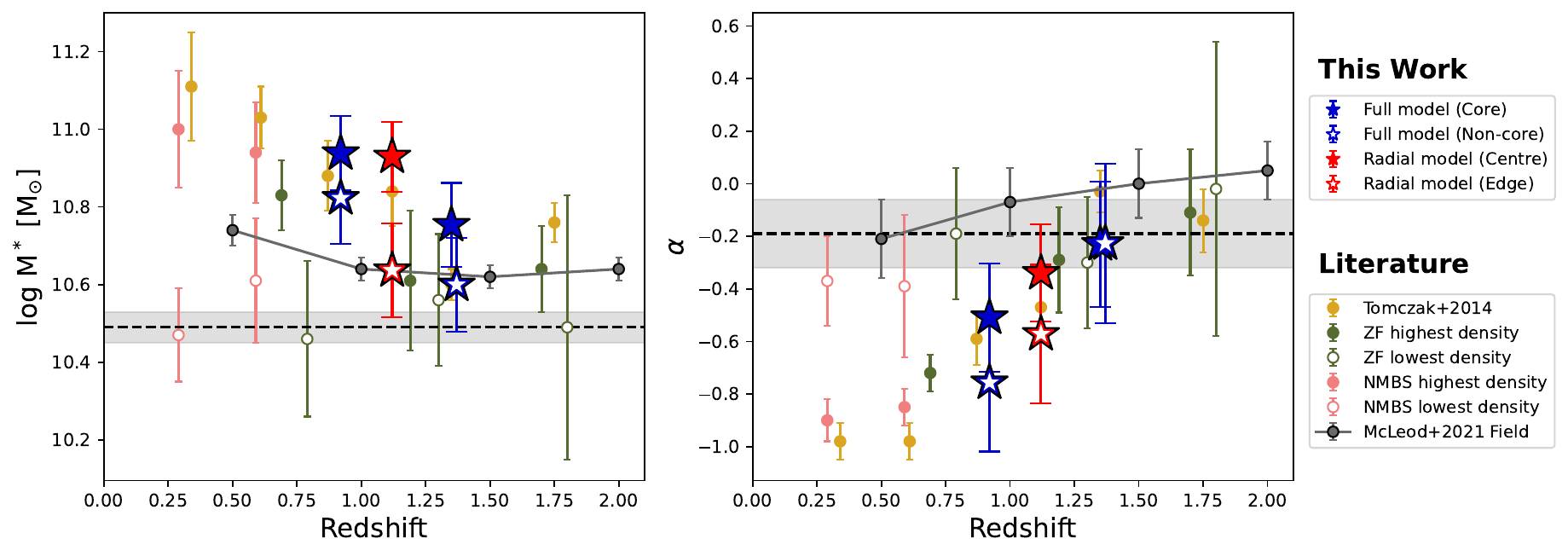}
\caption{Best-fit values for quiescent $M^*$ (\emph{left}) and $\alpha$ (\emph{right}) parameters in various environments through redshift. This figure is an adaptation of Figure 4 in \citet{Papovich2018}, and shows their best-fits from ZFOURGE, NMBS, and the related \citet{Tomczak2014} results. We compare our full parameter model results (blue stars) from the four edge populations in Figure \ref{fig:Red_SMFs} (the same used in Figure \ref{fig:Q_params}). The red stars show the radial model results for the two populations used in Figure \ref{fig:Q_radial_model}, with those points placed at the median redshift of the sample. While all the data points generally follow the trend of \emph{open} points for lower density, and \emph{closed} points for higher density, our `Non-core/Edge' sample represents an environment more similar to the other `highest density' samples in the figure. The dark grey points show the quiescent field parameter values from \citet{McLeod2021} and the grey shaded regions represent the mean range of parameter values for the `lowest density' environments.}
\label{fig:Papo_full_model}
\end{figure*}

In Figure \ref{fig:Papo_full_model} we adapt and add the results from both our full model and the radial model results to Fig. 4 from \citet{Papovich2018}, where they show their fits of Schechter parameters for low- and high-density regimes of quiescent galaxies. Figure \ref{fig:Papo_full_model} shows that our fit values follow their total and highest density results very well.  Notably, both the $\alpha$ and $M^*$ values of our sample are consistent with their `field' values at the high-redshift bins of our sample, while both differ significantly from the field at the low-redshift bins. Figure \ref{fig:Papo_full_model} also shows that the core/centre have $\alpha$ values more consistent with the field, while the non-core/edge have $M^*$ values that are more consistent with the field. We explore the physical interpretation of this in \S~\ref{sec:Toy_Models}.

\subsection{Simple Quenching Pathway Model}\label{sec:Toy_Models}
The results discussed in \S~\ref{sec:Literature} support the idea of two separate channels for increasing the quenched fraction in $z\sim1$ clusters, as has been suggested in the literature \citep{Poggianti2006,Wild2016,Socolovsky2018,Remco2020,Webb2020,Baxter2023}. At the cluster edge, a mildly enhanced quenched fraction appears qualitatively consistent with expectations from a mass-independent environmental-quenching of infalling star-forming galaxies, leading to a steeper $\alpha$ for the quenched population. On the other hand, the centre appears to have a quiescent $\alpha$ value more similar to the field, but with a much larger quenched fraction, and notably a highly elevated characteristic mass that is significantly different from the field.

Similar to the experiment in \cite{Papovich2018}, we consider a simple model which considers the cluster quiescent SMF to be a combination of the field quiescent SMF and the field star-forming SMF. In their work they apply various mass-dependent and independent quenching profiles to the star-forming SMF in order to explore the efficacy of the mass-independent environmental-quenching model \citep{Peng2010,Kovac2014}. In our model, we aim to explore the relative importance between the environmental quenching model from \citet{Peng2010} and the `early mass-quenching' model proposed in \citetalias{Remco2020}. We parameterise the importance of the early mass-quenching model with the coefficient $A_Q$ applied to the field quiescent SMF, and the importance of the environmental-quenching model with the coefficient $B_{SF}$ applied to the field star-forming SMF. The functional form of our model is given as
\begin{equation}\label{eq:Toy_model}
    \phi(M)_{Q,C}=A_Q\phi(M)_{Q,F}+B_{SF}\phi(M)_{SF,F}\;\;,
\end{equation}
where $\phi(M)_{Q,C}$ is the cluster quiescent population, $\phi(M)_{Q,F}$ is the quiescent field population, and $\phi(M)_{SF,F}$ is the star-forming field population. The overall normalisation of this combination is set by requiring the integral of $\phi(M)_{Q,C}$ to be equal to that of the observed data (or more specifically the parameterised fit). This ensures the total quenched fraction of our model matches that of the data, and means there is only one free parameter in our model (either $A_Q$ or $B_{SF}$).

\begin{figure*}
\centering
\includegraphics[width=1.0\linewidth]{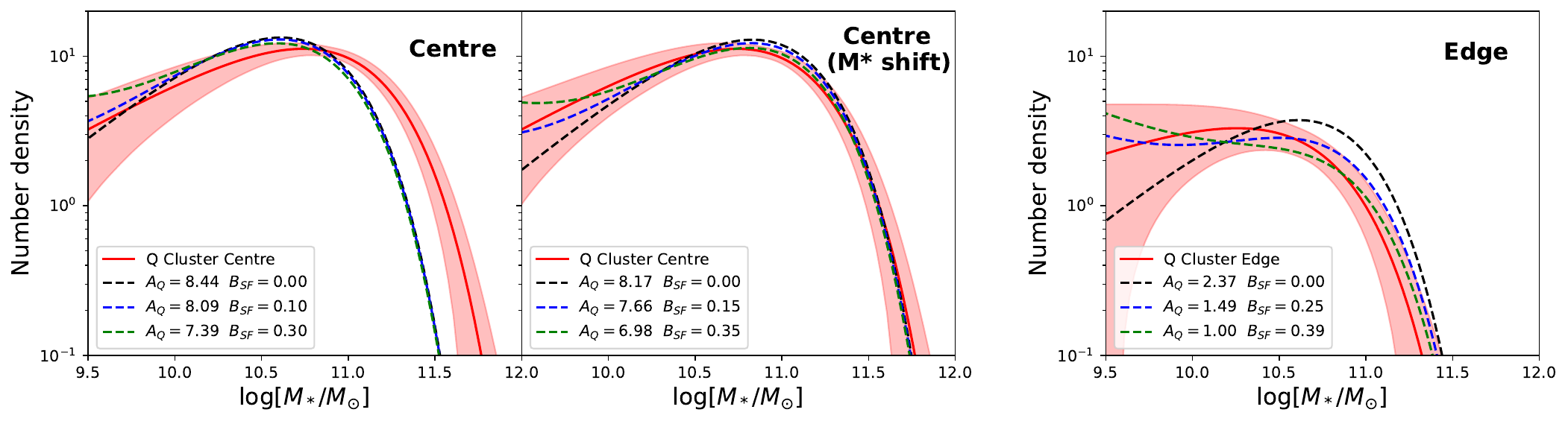}
\caption{A simple experiment to observe the significance of two independent quenching pathways in different $z\sim1$ cluster environments. The first component (enhanced mass quenching) is represented by a multiple of the field quiescent SMF, and the second (environmental quenching) is represented by a mass-independent fraction of the field star-forming SMF (best-fit values for both from \citealt{McLeod2021}). We quantify the relative importance of each pathway using the normalisation coefficient $A_Q$ and $B_{SF}$, respectively (see Equation \ref{eq:Toy_model}). The coloured dashed lines show different coefficient combinations explored to replicate the observed quiescent cluster population in each subplot (solid red curve with 1-$\sigma$ errors), with their coefficient values indicated in the legends. All combinations are constrained so the total SMF has the same integral as the cluster SMF within our sample stellar mass range ($M_*>10^{9.5}M_{\odot}$). \emph{Left}: Various combinations to replicate the quiescent cluster centre SMF. None of the models are able to fully describe the data. \emph{Middle}: Same as the left panel, but the $M^*$ value of the enhanced mass quenching population has been increased by 0.24 dex (see \S~\ref{sec:Toy_Models}). In this case the model provides a good match to the data. It requires a large enhancement relative to the field ($A_Q\gg 1$), yet still allows for significant levels of environment quenching (up to $B_{SF}\approx$ 0.15 - 0.35). \emph{Right}: Various combinations to replicate the quiescent cluster edge SMF. This population is well represented without any enhancement in mass-quenching ($A_Q\approx 1$), and allows a significant amount of environmental quenching ($B_{SF}\lesssim 0.4$).}
\label{fig:Toy_models}
\end{figure*}

The left panel of Figure \ref{fig:Toy_models} shows the SMF of the quiescent cluster centre and three different model combinations, each with increasing levels of environmental-quenching ($B_{SF}$ = 0.0, 0.1, 0.3). With the centre having an $\alpha$ value more consistent with the quiescent field value (see Figure \ref{fig:Q_radial_model}), it is unsurprising that any major addition of an environmental-quenching component will lead to a low-mass slope that diverges from the values observed in the cluster centre. However, even in the case where the model is dominated entirely through early mass-quenching ($B_{SF}$ = 0.0), the model is unable to reproduce the high-mass quiescent population, which is due to the larger $M^*$ value. To account for this, we consider models where $M^*$ is larger for the mass-quenched field population that ends up in the cluster. This is shown in the middle panel of Figure \ref{fig:Toy_models}, where the mass-quenched field $M^*$ is increased by a factor of 0.24 dex, which is approximately the increase we see in the centre from the field values (Figure \ref{fig:Q_radial_model}). With this change we are able to reproduce the observed SMF well.  Moreover, models that include significant levels of environmental-quenching (up to $B_{SF}\approx$ 0.15 - 0.35) are still consistent with the data. In both of these figures, the value of $A_Q$ is much higher than unity. This reflects the much higher quenched fraction in the cluster cores, and further shows the quiescent centre SMF has a similar shape to the field, but with a higher normalisation. This appears consistent with the interpretation of \citetalias{Remco2020} and \citet{Webb2020}, that the cluster core is dominated by quiescent galaxies that quenched similarly to the field, but with greater efficiency.

The right panel of Figure \ref{fig:Toy_models} shows the equivalent experiment (without the increase in field $M^*$) but evaluated for the cluster edge region. For this population we find that without an environmental-quenching component, the shape of the predicted SMF does not reproduce the observed trend. In the other two model combinations we decrease the early mass-quenching component, which allows for increased levels of environmental-quenching. The model that includes only environmental-quenching ($A_Q = 1.0$, $B_{SF} = 0.39$), does the best at reproducing both the low-mass and high-mass regimes of the observed SMF. This indicates that the edge region is dominated by environmental-quenching, and requires a minimal, if any, early mass-quenching component.

\subsection{Interpretation}
By considering the cluster populations as a function of cluster-centric radius, our work provides a way to reconcile some of the apparently discrepant results in the literature for the early evolution of cluster galaxies. While the outer regions are dominated by galaxies that quenched recently ($z\lesssim 3)$ during cluster assembly \citep[e.g.][]{Baxter2022,Werner2022,Edward2024}, the galaxies in the core were quenched much earlier, in the protocluster phase, via an accelerated mass-quenching mechanism \citep[e.g.][]{Remco2020,Webb2020}.

The origin of massive quenched field galaxies (those produced through `mass-quenching') is still a matter of ongoing research, but a majority are likely produced through AGN feedback \citep{Croton2006,Hopkins2008,Fabian2012,Cicone2014,Bremer2018}, potentially fed by rich gas mergers \citep{Kocevski2012,Fan2016,Mechtley2016,Shah2020,Xie2024}. A natural interpretation of the enhanced quiescent population in our cluster centre could therefore be an enhanced efficiency of merger-driven AGN feedback, as suggested by \citet{Pontzen2017} and \citet{Davies2022}. \citet{Tomczak2017} found a massive quiescent galaxy excess within the highest density regimes of their sample (ZFOURGE) just as we did, and explored models of galaxy-galaxy merging to help explain the discrepancy. Their merger model increased both the $\alpha$ and $M^*$ values of their fits, allowing for better fits of their highest density populations. Results from \citet{Rudnick2012} showed that as many as four mergers per quiescent galaxy are necessary for the quiescent SMF shape of $z=1.6$ clusters to match that of $z\approx0.6$ clusters (\citet{Rudnick2009}).

The plausibility of mergers playing a role alongside early mass-quenching increases with the evidence that the massive galaxies in the GOGREEN/GCLASS clusters were self-quenched or pre-processed prior to infall \citep{Webb2020,Werner2022}. The progenitors of cluster galaxies form within protoclusters, in which the most massive galaxies are approximately twice as massive as those in comparably-selected field samples \citep{Steidel2005,Hatch2011,Koyama2013,Cooke2014}. When measured explicitly, protocluster SMFs have enhanced $M^*$ values, with an increase of 0.2-0.4 dex from the field fits \citep{Cooke2014,Edward2024}, which is consistent with the increase of 0.24 dex we find necessary in our quiescent core model. \textcolor{black}{Some protocluster observations have also shown strong AGN fraction enhancements, in addition to that caused by their overdensity of high-mass galaxies \citep{Lehmer2013,Krishnan2017,Espinoza2024,Vito2024}. This is consistent with some evidence that the AGN fraction in clusters increases more rapidly with redshift than the field \citep[e.g.][]{Martini2013,Bufunda2017,Hashiguchi2023}}. Combined with moderate velocity dispersions that allow for significantly enhanced merger rates \citep{Lotz2013,Giddings2025}, the conditions that protoclusters provide could explain both the mechanisms that allow for greater efficiency of early mass-quenching, as well as an increased presence of high-mass quenched galaxies.

There is the caveat, however, that the early mass-quenching  mechanisms may not be happening in protoclusters at all. Simulation results indicate protoclusters have equivalent star formation rates as the field at $z>3$, but peak at earlier times \citep{Chiang2017,Muldrew2018}. \citet{Muldrew2018} note that this earlier peak is due to star formation being suppressed in protoclusters at $z<3$, rather than being enhanced at earlier times. Additional simulation results from \citet{Ahad2024} find that protoclusters have a biased population of infall galaxies, such that many of the galaxies supposed to be quenched by the protocluster could instead be explained through ``pure'' secular processes. Nevertheless, if this were the case the outcome would remain the same, with the cluster core population still being composed of these massive quenched galaxies, regardless of their exact quenching locations.

The cluster edge region follows a quenching model more similar to that described in \citet{Peng2010}. This might be expected as these galaxies are likely to originate in less-biased regions of the Universe, where the progenitor population may be more representative of the low-density field population. The later buildup of the deep gravitational potential and ICM then creates an environment that allows for the typical mass-independent environmental-quenching processes (e.g. ram-pressure stripping, strangulation, interaction-driven star formation or AGN activity) to become more efficient.  Some of these mechanisms may also operate in the overdense infall regions of these growing systems \citep[e.g.][]{Haines+12,Haines2015,Koulouridis+19,Koulouridis+24}. The efficiencies of these processes are proportional to the local density of the ICM \citep[or more explicitly $t_q\propto\rho^{-1/2}$, ][]{McGee2014}, so it is also expected that these processes would be even more efficient in the cluster centre. Figure \ref{fig:Toy_models} shows that, even though the centre is dominated by enhanced mass-quenching, it is still consistent with significant environment quenching of $B_{SF}\lesssim 0.35$, comparable to the constraints in the edge regions.

\subsection{Considerations and Future Work}
It is common to include Eddington bias \citep{Eddington1913} into SMF modelling to account for stellar mass uncertainties, which can significantly alter the shape of the SMF. This was neglected in our SMF fits in the present work because it requires a good understanding of mass uncertainties as a function of mass, redshift and galaxy type. If included, this would potentially increase the size of the posterior uncertainties, and change the fit values for $\alpha$ and $M^*$. However, we expect the dependence of $M^*$ on environment that we find would not be strongly affected, as long as the mass uncertainties are not themselves strongly environment-dependent.

The large variation in the global quenched fraction of the cluster sample as a function of redshift and cluster velocity dispersion indicates there are likely other dependencies that are primarily driving this scatter. The model in this work is robust and efficient enough to explore additional relationships and parameterisations of the evolution of cluster SMFs, including parameters related to the dynamical state of clusters. However, this implementation is currently limited by the sample size. This limitation will soon be mitigated with upcoming large-field surveys such as Euclid and LSST, which have the potential to observe local clusters through to high-\emph{z} protoclusters \citep{Brough2020,Euclid2025}. These surveys will primarily rely on photometric redshifts, and deep, wide-field spectroscopy will be required to assess cluster membership the way we have done for the GOGREEN and GCLASS clusters. Two examples of complementary surveys are CHANCES and MOONRISE. CHANCES will use the 4MOST telescope to obtain the spectra for all $\log{[M/M_\odot]}>10.0$ members of 50 galaxy clusters at $0.07<z<0.45$ \citep{Sifon2024}, and MOONRISE will use the MOONS instrument on the VLT to obtain spectra for potentially half a million galaxies at $0.9<z<2.6$ \citep{Maiolino2020}.

Throughout this work, it is apparent that the emergence of the conventional environmentally-quenched population occurs primarily at low stellar masses, and is reflected in the  $\alpha$ parameter.  This parameter is still poorly constrained in our fits, due to the limited depth of the imaging and the sample size. This demonstrates the necessity of acquiring even deeper photometry for clusters in any future samples \citep[see also][]{Gully2025}. For example, in field quiescent galaxies, an upturn in the SMF (often interpreted as due to environmental quenching) is only observed at $M_*\lesssim10^{9.5}M_\odot$\citep{McLeod2021}.

A major conclusion of this work is that protoclusters might have a significant effect on shaping the SMFs of clusters within the redshift range of our sample.  Extending this analysis to $z>2$ is necessary to observe the buildup of massive quenched galaxies which become the progenitors of the cluster core galaxies, and in turn explore the potential mechanisms that are causing them to quench. Further observations of increased merger rates as well as enhanced AGN fractions would help with the understanding of the results from this work. This would add to the current body of previous works, which have found evidence of elevated mass quenched populations within protoclusters \citep{Muldrew2018,Hill2022,McConachie2022,Edward2024,Tanaka2024}.

\section{Conclusions}\label{sec:Conclusions}
In this work we measure the SMFs down to a mass limit of $\log{(M/M_\odot)}=9.5$ for seventeen galaxy clusters in the redshift range of $0.8<z<1.5$ from the GOGREEN and GCLASS surveys. We parameterise the $\alpha$ and $M^*$ parameters of the star-forming and quiescent galaxies to be themselves functions of cluster-centric radius and redshift, and parameterise the total quenched fraction within $R_{200}$ to be a function of redshift and cluster velocity dispersion. We compare our results with a field sample, and model the contribution of environmental quenching models for the quiescent populations at two radial regions. Our main findings are:

\begin{itemize}
  \item[$\bullet$] The cluster quenched fraction depends significantly on both stellar mass and cluster-centric radius. It increases strongly with galaxy stellar mass and proximity to the cluster core, and is elevated from the field values \citep{McLeod2021} for all mass and radii, with the core having highly elevated values (Figures \ref{fig:Fq_M} and \ref{fig:Fq_R}). This is consistent with the results previously found with subsets of the same data \citep{Muzzin2012,Remco2020,Baxter2023}.
  \item[$\bullet$] There is a weak, low-significance trend of decreasing quenched fraction with redshift over $0.8<z<1.5$ (Figure \ref{fig:Fq_R}), and no dependence of quenched fraction on cluster velocity dispersion. There is a significant cluster-to-cluster variation in the quenched fraction beyond any correlation with  these variables.
  \item[$\bullet$] The stellar mass functions of the star-forming galaxies in our clusters show no radial or redshift dependence, and are consistent with field values (Figure \ref{fig:SF_params}).
  \item[$\bullet$] The stellar mass functions of the quiescent galaxies in our sample show no significant evidence for redshift dependence, and $\sim 2\sigma$ evidence for radial dependence (Figures \ref{fig:Q_params} and \ref{fig:Q_radial_model}).  The Bayesian Information Criterion indicates that a model in which $\alpha$ or $M^\ast$ vary with radius is preferred with moderate significance over one in which they are both constant. It also indicates a more complex model does not significantly improve the fit.
  \item[$\bullet$] The cluster centre has a high quenched fraction ($\sim 70$ per cent) and a quiescent galaxy SMF that is similar in shape to that of the quiescent field, but with a $M^\ast$ parameter that is $\sim 0.24$ dex larger. This is indicative of the mass-dependent `early mass-quenching' model proposed by \citet{Remco2020}, in which protocluster galaxies are quenched in a manner similar to field galaxies, but at an accelerated rate.
  \item[$\bullet$] At the edge of the cluster, the elevated quenched fraction relative to the field is instead due primarily to an excess of low-mass quiescent galaxies, leading to the steeper (more negative) $\alpha$, characteristic of mass-independent "environment-quenching models" \citep[e.g.][]{Peng2010}.  
  \item[$\bullet$] With a simple toy model (Figure \ref{fig:Toy_models}) we show that the quiescent SMF throughout the cluster is consistent with up to $\sim 40$ per cent contribution from environment quenching.  Deeper data are required to improve constraints on $\alpha$ and establish this fraction with more precision.
\end{itemize}

In conclusion, the high quenched fraction relative to the field in these $z\sim 1$ clusters is dominated by the core population, which appears to have formed through an accelerated mass-quenching mechanism similar to what occurs for central galaxies.  However, we cannot rule out a significant ($\lesssim 40$ per cent) contribution from environmental quenching. Future studies based on large, wide-field surveys (e.g. from Euclid and LSST), and combined with accurate redshift measurements (e.g. CHANCES and MOONRISE), should be able to probe the transition of dominant quenching models from protoclusters to massive low-redshift clusters.

\section*{Acknowledgements}
This research is primarily supported by an NSERC Discovery Grant to MLB.  DCB is supported by an NSF Astronomy and Astrophysics Postdoctoral Fellowship under award AST-2303800, and by the UC Chancellor's Postdoctoral Fellowship. GC acknowledges the support from the Next Generation EU funds within the National Recovery and Resilience Plan (PNRR). RD gratefully acknowledges support by the ANID BASAL project FB210003. GW gratefully acknowledges support from the National Science Foundation through grant AST-2205189. YMB acknowledges support from UK Research and Innovation through a Future Leaders Fellowship (grant agreement MR/X035166/1) and financial support from the Swiss National Science Foundation (SNSF) under project 200021\_213076. 

\textit{Software}: {\sc Astropy} \citep{Astropy2022}, {\sc Corner} \citep{corner}, {\sc Emcee} \citep{emcee}, {\sc GalSim} \citep{galsim}, {\sc Matplotlib} \citep{matplotlib}, {\sc Numpy} \citep{Numpy2011}, {\sc Pandas} \citep{pandas}, {\sc SExtractor} \citep{sextractor}, {\sc SciPy} \citep{Scipy}.

\section*{Data Availability}
Data used in this work were obtained from the GOGREEN and GCLASS Data Release catalogues, which can be found on \href{https://apps.canfar.net/storage/vault/list/GOGREEN/}{CANFAR}, and are described in \citet{Balogh2021}.



\bibliographystyle{mnras}
\bibliography{ms} 



\appendix

\section{Magnitude Limit and Completeness}\label{appex:MagCompl}
\begin{figure}
\centering
\includegraphics[width=0.99\linewidth]{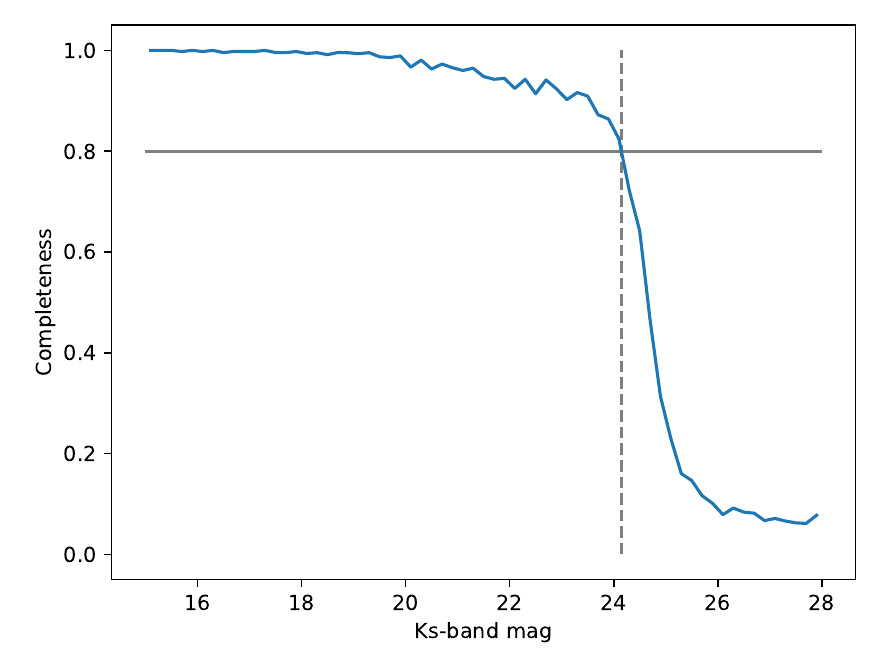}
\caption{Recovery rate of injected mock sources against their intrinsic magnitudes. The limiting magnitude is set as the value where the recovery rate is 80 per cent (solid grey line), with the value of 24.15 mag shown with the dashed grey line. This form is the ``naive completeness'', without the inclusion of magnitude bias and Eddington bias.}
\label{fig:completeness}
\end{figure}

As mentioned in Sec \ref{sec:1033}, we followed the same procedure for SpARCS-1033 as the other clusters in \citetalias{Remco2020}, where the limiting magnitude is based on the recovery rate of mock sources that are injected on the raw $K_s$-band image. We created the mock sources using the \texttt{GALSIM} software \citep{galsim}, with all the sources having the same S\'ersic = 1 light profile. The rest of the parameters are randomly drawn from uniform distributions, with the magnitudes in the range of 15-28, the half-light radii in the range of 1-3 kpc, the ellipticities in the range of 0.0-0.2, and all of random orientation. A total of 30,000 mock sources were injected across 60 sets of the raw image, so not to alter the underlying characteristics of the original image. The sources are recovered using \texttt{SExtractor} \citep{sextractor}, and the individual recoveries of each run are stacked to give the overall completeness curve, as seen in Figure \ref{fig:completeness}. 

\begin{figure}
\centering
\includegraphics[width=0.99\linewidth]{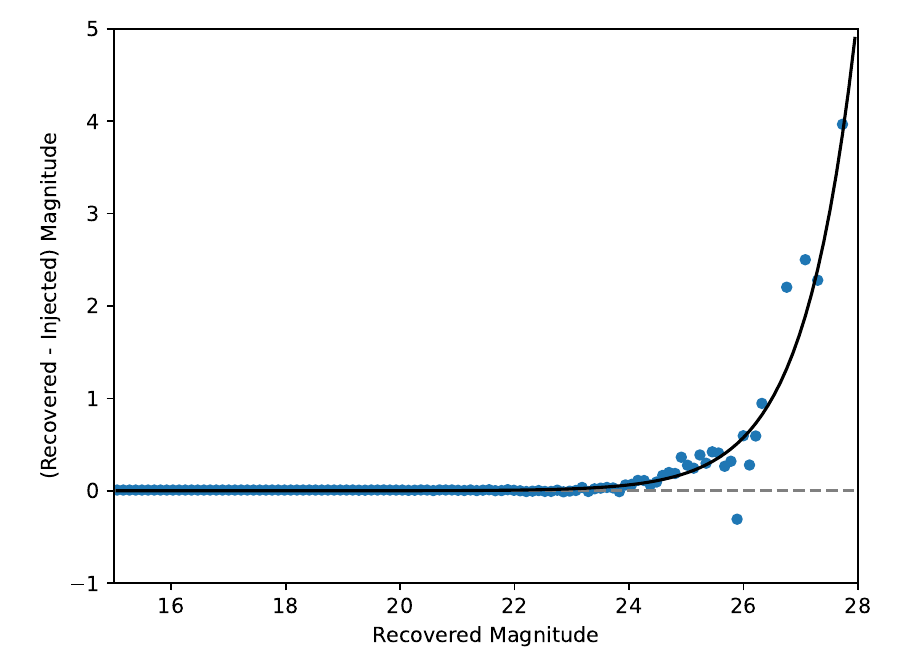}
\caption{Binned difference between the recovered and injected magnitudes of the injected mock samples against their recovered magnitudes. The fit shows the tendency for \texttt{SExtractor} to measure sources as fainter than their true values, especially at higher magnitudes.}
\label{fig:magbias}
\end{figure}

Following the definition in \citetalias{Remco2020}, the limiting magnitude is set where the mock source recovery rate is 80 per cent. For the curve shown in Figure \ref{fig:completeness}, this corresponds to a magnitude of 24.15 mag$_{\text{AB}}$. This curve shows the recovery of intrinsic magnitudes, and so is missing two main sources of bias. The first is a magnitude bias, which corrects for the discrepancy between the recovered and intrinsic magnitudes of the mock sources, with the binned points and fit shown in Figure \ref{fig:magbias}. The second is Eddington bias, which affects the completeness distribution, especially the steep slope at higher magnitudes, with the random magnitude scatter resulting in more fainter sources scattering into the bins of brighter sources \citep{Eddington1913}. The correction factors for both biases were matched with those in \citetalias{Remco2020}, leading to a minor correction and giving a limiting magnitude value of 23.89 mag$_{\text{AB}}$ for SpARCS-1033.

\begin{figure}
\centering
\includegraphics[width=0.99\linewidth]{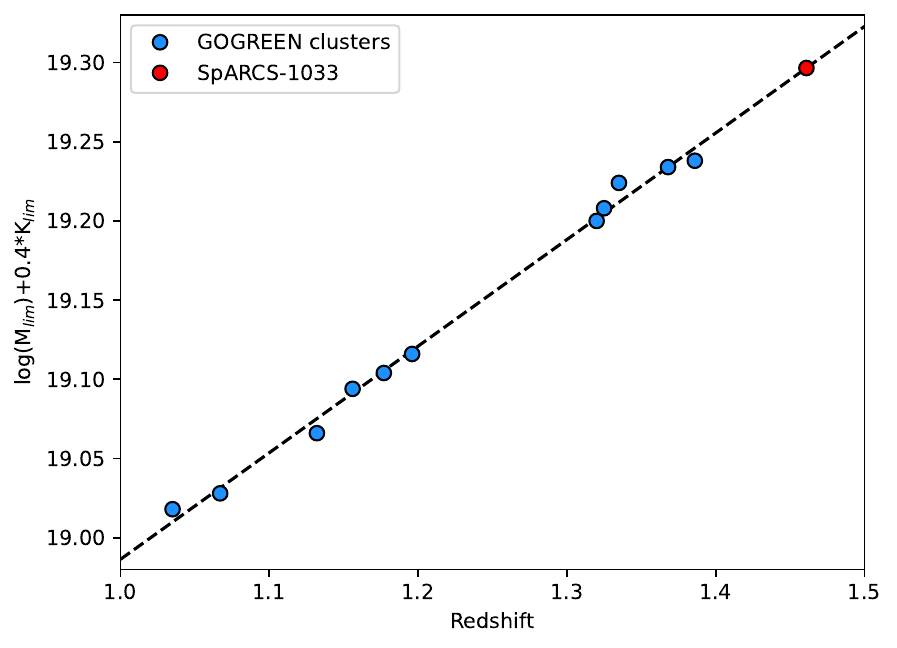}
\caption{Relation to approximate the stellar mass for the 11 GOGREEN clusters given in \citetalias{Remco2020}, plotted against the redshift of each cluster. The black dashed fit line allows us to extend the relation to the redshift of SpARCS-1033, and solve for its limiting stellar mass.}
\label{fig:masstolight}
\end{figure}

Using the corrected magnitude limit we can calculate the stellar mass limit of the cluster. Given the small range of redshift and limiting masses of the cluster sample, we can estimate the mass limit of SpARCS-1033 via a simple extrapolation of the full \texttt{FAST}-derived fits completed for the other clusters in \citetalias{Remco2020}. Using the stellar mass and magnitude limits, we can create a redshift dependent relation that approximates the cluster's mass-to-light ratios ($M_*/L$) of the form $\log M_{lim}+0.4K_{s,lim}=\log \left(\frac{M_*}{L}\right)+\log(4\pi)+2\log\left(D_L\right)$. By plotting this relation for each GOGREEN cluster through their redshift, as shown in Figure \ref{fig:masstolight}, we are able to fit this trend and extend to the redshift of SpARCS-1033 and solve for its limiting stellar mass of 9.74 $\pm$ 0.03 $\log{M_{\odot}}$ (with all cluster values stated in Table \ref{tab:Cluster_values}).

\section{Quenched Fraction Trends and Corner Plot}\label{app-B}
\begin{figure}
\centering
\includegraphics[width=1.0\linewidth]{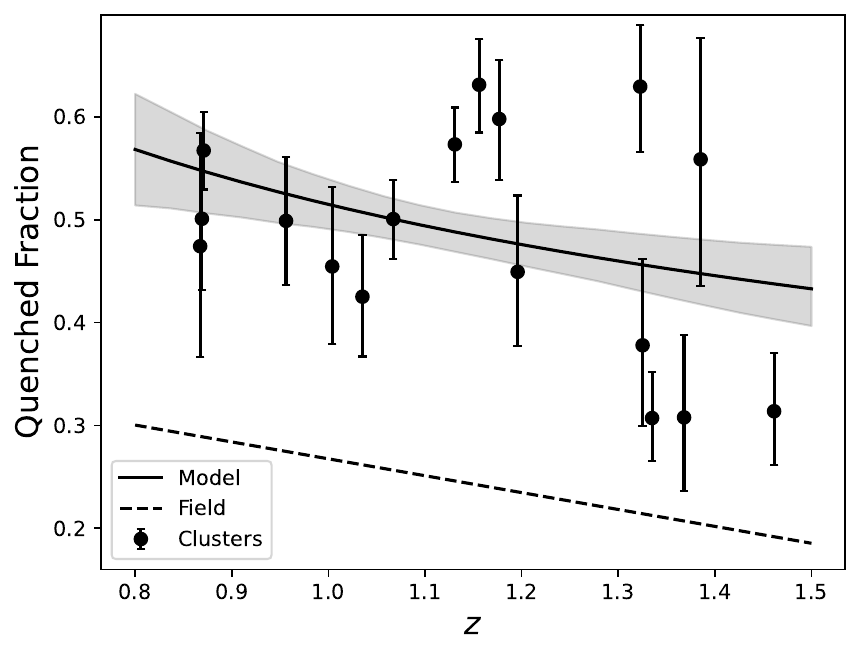}
\caption{The fit quenched fraction as a function of redshift, evaluated on the mean cluster velocity dispersion of the sample ($\sigma = 891$ km/s). The errors are the 1-$\sigma$ parameter uncertainties. The binned points show the quenched fraction of the individual 17 clusters in the sample and their corresponding binomial confidence intervals. The field reference corresponds to the fit quenched fraction function from \citet{McLeod2021}.}
\label{fig:Fq_z}
\end{figure}

The weak redshift evolution of the quenched fraction indicated in Figure~\ref{fig:Fq_R} is shown explicitly in Figure \ref{fig:Fq_z},  compared with the field from \citet{McLeod2021}. The fits to our data show the cluster quenched fraction, integrated over all radii and stellar masses, is significantly higher than the field at all redshifts, by about 25 per cent.  This result is also consistent with the previous literature \citep[e.g.][]{Peng2010,Muzzin2012,Lee2015,Balogh2016,Kawinwanichakij2016,Paccagnella2016,Papovich2018,Pintos-Castro2019}, including work on this sample \citep{Baxter2022}. Also shown is the binned quenched fraction values of the individual clusters. The dispersion they exhibit is large compared to the trend with redshift, which indicates that redshift is not the dominant factor driving cluster-to-cluster variations. In addition, the significant scatter is fully captured by the uncertainty in the model.

\begin{figure}
\centering
\includegraphics[width=1.0\linewidth]{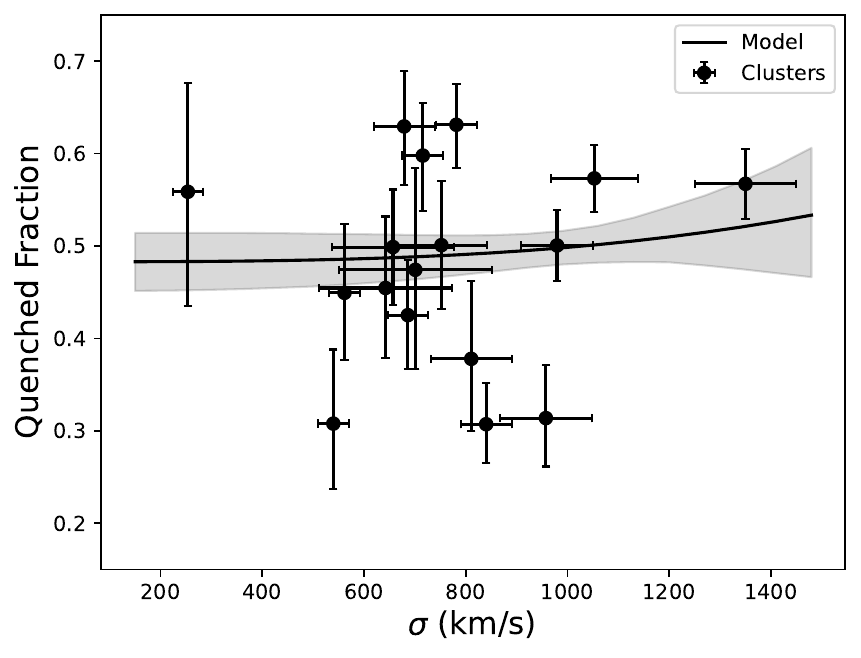}
\caption{The modelled quenched fraction as a function of cluster velocity dispersion, including the $\pm1$-$\sigma$ parameter uncertainties. The binned quenched fraction of the individual clusters is also plotted, with the vertical errors being their confidence intervals, and their uncertainties in velocity dispersion as their horizontal errors, matching those given in \citet{Balogh2021}.}
\label{fig:Fq_sig}
\end{figure}

\begin{figure}
\centering
\includegraphics[width=1.0\linewidth]{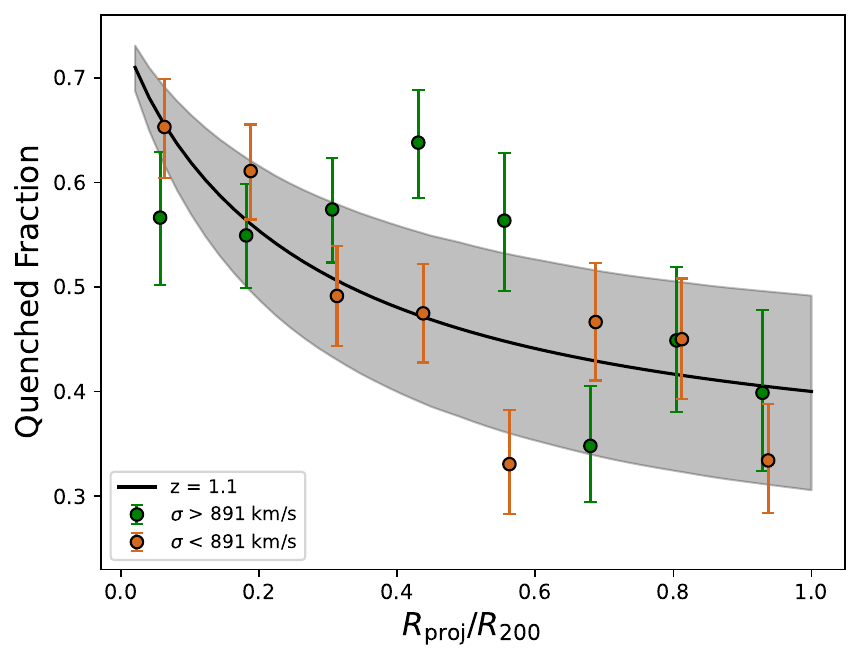}
\caption{Binned sample points and the fit functional form of the quenched fraction as a function of cluster-centric radius. The sample is split into two velocity dispersion bins on the mean velocity dispersion of the sample ($\sigma\simeq 891$ km/s). The function is evaluated at the mean redshift of the cluster ($z\simeq1.1$), and shows the 1-$\sigma$ parameter uncertainties. The errors on the binned points are their binomial confidence intervals.}
\label{fig:Fq_sigma}
\end{figure}

We also consider a potential dependence of quenched fraction on halo mass. We use the velocity dispersion values (see \S~\ref{sec:Radius_Likelihood}) as a proxy for the dynamical mass of the clusters. Figure \ref{fig:Fq_sig} shows the quenched fraction as a function of velocity dispersion, and like Figure \ref{fig:Fq_z}, shows the relationship against the quenched fraction of the individually binned clusters. We find no significant dependence of global quenched fraction on cluster velocity dispersion, and again the large variation on a cluster-to-cluster basis indicates velocity dispersion is not the main parameter to describe these differences. We also found no significant difference in the quenched fraction radial trends of clusters based on their velocity dispersion, as shown in Figure \ref{fig:Fq_sigma}.

\begin{figure*}
\centering
\includegraphics[width=1.0\linewidth]{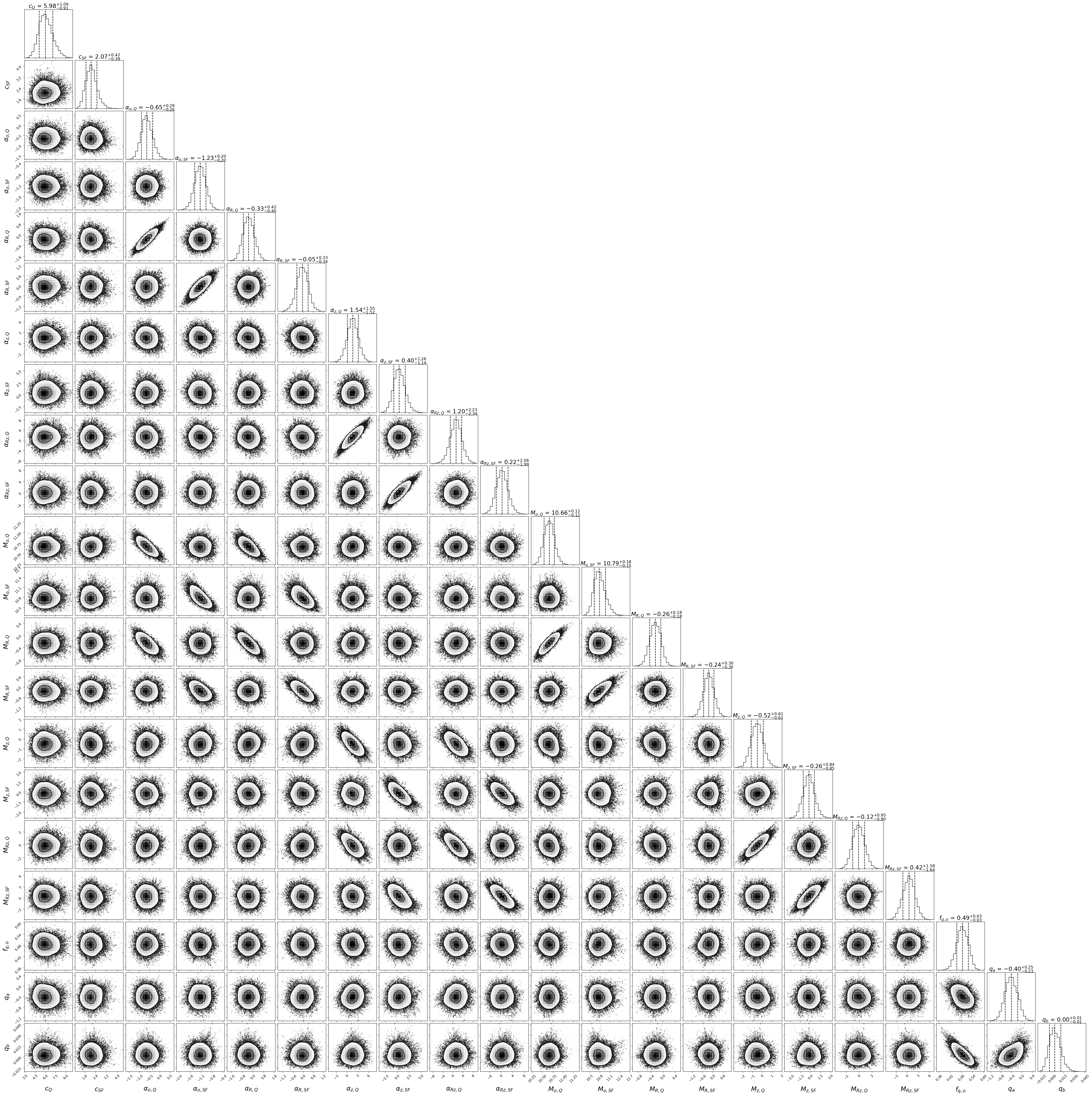}
\caption{Complete corner plot of the full model sample fit, corresponding to the values in Table \ref{tab:params}. The 16, 50, and 84 percentiles (the 1-$\sigma$ and midpoint values) are shown with the vertical dashed lines in the 1-D histograms, and the 2-D histograms show the 1-$\sigma$ and 2-$\sigma$ contours. The plot was constructed using the \texttt{corner} software \citep{corner}.}
\label{fig:full_corner}
\end{figure*}


\bsp	
\label{lastpage}
\end{document}